\documentclass[12pt,a4paper,twoside]{article}

\usepackage[dvips]{graphics}
\usepackage{cite}

\newcommand{\costt}  {\ensuremath{\cos\theta_T}}
\newcommand{\qqbar}  {\ensuremath{\mathrm{q\overline{q}}}}

\newcommand{\epem}   {\ensuremath{\mathrm{e^+e^-}}}

\newcommand{\as}     {\ensuremath{\alpha_s}}

\newcommand{\asq}    {\ensuremath{\alpha_s(Q)}}
\newcommand{\asmz}   {\ensuremath{\alpha_s(M_{\mathrm{Z^0}})}}

\newcommand{\oaa}    {\ensuremath{\mathcal{O}(\alpha_s^2)}}
\newcommand{\oaaa}   {\ensuremath{\mathcal{O}(\alpha_s^3)}}
\newcommand{\znull}  {\ensuremath{\mathrm{Z^0}}}
\newcommand{\zzero}  {\ensuremath{\mathrm{Z^0}}}
\newcommand{\zg}     {\ensuremath{(\mathrm{Z^0}/\gamma)^{*}}}
\newcommand{\mz}     {\ensuremath{M_{\mathrm{Z^0}}}}

\newcommand{\bt}     {\ensuremath{B_T}}
\newcommand{\bw}     {\ensuremath{B_W}}
\newcommand{\cp}     {\ensuremath{C}}
\newcommand{\mh}     {\ensuremath{M_H}}
\newcommand{\thr}    {\ensuremath{1-T}}
\newcommand{\tma}    {\ensuremath{T_{\mathrm{major}}}}
\newcommand{\tmi}    {\ensuremath{T_{\mathrm{minor}}}}

\newcommand{\chisqd} {\ensuremath{\chi^2/\mathrm{d.o.f.}}}
\newcommand{\xmu}    {\ensuremath{x_{\mu}}}
\newcommand{\xp}     {\ensuremath{x_p}}
\newcommand{\yp}     {\ensuremath{y}}
\newcommand{\ksip}   {\ensuremath{\xi_p}}
\newcommand{\ksimed} {\ensuremath{\xi_m}}
\newcommand{\ksinul} {\ensuremath{\xi_0}}
\newcommand{\ksimean}{\ensuremath{\langle\xi_p\rangle}}
\newcommand{\ptin}   {\ensuremath{p_{\perp}^{\mathrm{in}}}}
\newcommand{\ptout}  {\ensuremath{p_{\perp}^{\mathrm{out}}}}

\newcommand{\nch}    {\ensuremath{n_{\mathrm{ch}}}}
\newcommand{\mnch}   {\ensuremath{\langle n_{\mathrm{ch}}\rangle}}

\newcommand{\ycut}   {\ensuremath{y_{\mathrm{cut}}}}

\newcommand{\stat}   {\ensuremath{\mathrm{(stat.)}}}
\newcommand{\syst}   {\ensuremath{\mathrm{(syst.)}}}
\newcommand{\lnr}    {\ensuremath{\ln(R)}}
\newcommand{\wqcd}   {\ensuremath{W_{\mathrm{QCD}}}}
\newcommand{\ytwothree}   {\ensuremath{y^D_{23}}}
\newcommand{\rtwo}   {\ensuremath{{\cal R}_2}}
\newcommand{\ctwo}   {\ensuremath{C_2}}
\newcommand{\rs}     {\ensuremath{\sqrt{s}}}

\newcommand{\rsp}    {\ensuremath{\sqrt{s'}}}
\newcommand{\sprime} {\ensuremath{s'}}
\newcommand{\tautau} {\ensuremath{\tau^+\tau^-}}

\newcommand{\ww}     {\ensuremath{\mathrm{W^+W^-}}}
\newcommand{\zz}     {\ensuremath{\mathrm{Z^0Z^0}}}

\newcommand{\bm}[1]  {\mbox{\boldmath\ensuremath{#1}}}
\newcommand{\perc}   {\%}
\newcommand{\sdsd}   {\ensuremath{1/\sigma\cdot\mathrm{d}\sigma/\mathrm{d}}}
\newcommand{\sdscd}  {\ensuremath{1/\sigma\cdot\mathrm{d}\sigma_{\mathrm{ch}}
                                  /\mathrm{d}}}

\newcommand{\gev}    {\ensuremath{\mathrm{GeV}}}

\newcounter{hours}
\newcounter{minutes}

\newcommand{\Printtime}{%
  \setcounter{hours}{\time/60}%
  \setcounter{minutes}{\time-\value{hours}*60}%
  \ifthenelse{\value{hours}<10}{0}{}\thehours:%
  \ifthenelse{\value{minutes}<10}{0}{}\theminutes}

\begin{document}


%
\begin{titlepage}
\begin{center}{\large   EUROPEAN ORGANIZATION FOR NUCLEAR RESEARCH
}\end{center}\bigskip
\begin{flushright}
       CERN-EP/99-178   \\ December 17, 1999 \\ (revised February 1, 2000)
\end{flushright}
\bigskip\bigskip\bigskip\bigskip\bigskip
\begin{center}
{\huge\bf
QCD studies with \bm{\epem} annihilation data at 172-189 GeV
}
\end{center}
\bigskip\bigskip
\begin{center}
{\LARGE The OPAL Collaboration}
\end{center}
\bigskip\bigskip\bigskip
\begin{center}{\large  Abstract}\end{center}
We have studied hadronic events from \epem annihilation data at
centre-of-mass energies of $\rs=$172, 183 and 189~GeV. 
The total integrated luminosity
of the three samples, measured with the OPAL detector, corresponds to
250 pb$^{-1}$.

We present distributions of event shape variables, charged particle
multiplicity and momentum, measured separately in the three data
samples.  From these we extract measurements of the
strong coupling \as, the mean charged particle multiplicity \mnch\
and the peak position \ksinul\ in the $\ksip=\ln(1/\xp)$ distribution.

In general the data are described well by analytic QCD
calculations and Monte Carlo models. Our measured values of \as,
\mnch\ and \ksinul\ are consistent with previous determinations at
\rs=\mz.
\bigskip\bigskip\bigskip\bigskip
\bigskip
%
\bigskip
\begin{center}
{\large 
(Submitted to European Physical Journal C) }
\end{center}
 
%
\begin{center}
%





\end{center}

\end{titlepage}
\begin{center}{\Large        The OPAL Collaboration
}\end{center}\bigskip
\begin{center}{
G.\thinspace Abbiendi$^{  2}$,
K.\thinspace Ackerstaff$^{  8}$,
P.F.\thinspace Akesson$^{  3}$,
G.\thinspace Alexander$^{ 22}$,
J.\thinspace Allison$^{ 16}$,
K.J.\thinspace Anderson$^{  9}$,
S.\thinspace Arcelli$^{ 17}$,
S.\thinspace Asai$^{ 23}$,
S.F.\thinspace Ashby$^{  1}$,
D.\thinspace Axen$^{ 27}$
% ...
}\end{center}\bigskip
\bigskip
$^{  1}$School of Physics and Astronomy, University of Birmingham,
Birmingham B15 2TT, UK
\newline
$^{  2}$Dipartimento di Fisica dell' Universit\`a di Bologna and INFN,
I-40126 Bologna, Italy
\newline
$^{  3}$Physikalisches Institut, Universit\"at Bonn,
D-53115 Bonn, Germany
\newline
% ...
$^{  8}$CERN, European Organisation for Particle Physics,
CH-1211 Geneva 23, Switzerland
\newline
$^{  9}$Enrico Fermi Institute and Department of Physics,
University of Chicago, Chicago IL 60637, USA
\newline
% ...
$^{ 16}$Department of Physics, Schuster Laboratory, The University,
Manchester M13 9PL, UK
\newline
$^{ 17}$Department of Physics, University of Maryland,
College Park, MD 20742, USA
\newline
% ...
$^{ 22}$Department of Physics and Astronomy, Tel Aviv University,
Tel Aviv 69978, Israel
\newline
$^{ 23}$International Centre for Elementary Particle Physics and
Department of Physics, University of Tokyo, Tokyo 113-0033, and
Kobe University, Kobe 657-8501, Japan
\newline
% ...
$^{ 27}$University of British Columbia, Department of Physics,
Vancouver BC V6T 1Z1, Canada
\newline
% ...
%
%
\newpage
\section{Introduction}

We study the general features of hadronic decays in $\epem\rightarrow\zg\rightarrow\qqbar$
 reactions
at the highest available centre-of-mass (c.m.) energies, as a continuation of
our earlier publications at c.m.\ energies of $\rs=130-136$~GeV~\cite{OPALPR158}
and $\rs =161$~GeV~\cite{OPALPR197}. 

In the last three years LEP
produced \epem\ collisions at centre-of-mass energies of $\rs\approx 172$,
183 and 189~GeV. The total integrated luminosity measured with the OPAL
detector at these energies corresponds to approximately 250 pb$^{-1}$.

Previous studies using \epem\ annihilation data at c.m.\ energies up
to $\rs = 183$~GeV have shown that QCD based models and calculations
give a good description of the observed
data~\cite{OPALPR197,OPALPR158,alephas133,l3as133,delphias133,delphinch133,
l3as161172,l3as183,delphinch172,bethke/pilcher}.
Here we examine the overall consistency of QCD at
yet higher c.m.\ energies, with improved statistical precision
owing to the dramatic increase in integrated luminosity at
c.m.~energies above $\rs = 161$~GeV.

We determine the strong coupling strength \as\ at each of the three
energies.  The large data sample at 189~GeV, together with the fact
that the hadronization corrections become smaller at higher c.m.\
energies, allows a precision comparable with that achieved
at $\rs=\mz$ using event shapes. We compare our results to
those obtained at lower c.m.\ energies, notably at $\rs=\mz$.  We also
determine the charged particle distributions and study the evolution with the 
c.m.\ energy of the mean charged particle multiplicity and the peak position 
\ksinul\ in the $\ksip=\ln(1/\xp)$  distribution.
Jet-multiplicity related observables have been analized
and used to determine \as\ at these c.m.\ energies in a separate study of 
\epem\ annihilation data taken by the JADE and OPAL experiments
at c.m.\ energies between 35 and 189~GeV~\cite{opaljade}.


The majority of hadronic events produced at c.m.\ energies
above the \zzero\ resonance are `radiative' events in which initial
state photon radiation reduces the energy of the hadronic system
to about \mz.  An experimental separation between radiative and
non-radiative events is therefore required.  In addition, at energies
above 160~GeV, \ww\ production becomes kinematically possible
and contributes a significant background to the $\zg\rightarrow\qqbar$ events. For $\rs >
180$~GeV, \zz\ production must also be taken into
account.

In this paper we use similar techniques to those of our previous
analysis of \epem\ annihilation data at c.m.\ energies of 
130--136~GeV~\cite{OPALPR158} and 161~GeV~\cite{OPALPR197}.  In
Section~\ref{sec_detector} we briefly describe the OPAL detector, in
Section~\ref{sec_datamc} we describe the samples of data and simulated
events and in Section~\ref{sec_anal} we explain our selection
requirements and analysis procedures. In Section~\ref{sec_results} we
then present the results in the form of event shape variables,
determination of \as, charged particle multiplicities and 
momentum spectra. Section~\ref{sec_conc} gives a summary and
conclusion.

\section{ The OPAL detector}
\label{sec_detector}
The OPAL 
detector operates at the LEP \epem\ collider at CERN.  A detailed
description can be found in reference~\cite{opaltechnicalpaper}. The
analysis presented here relies mainly on the measurements of momenta
and directions of charged tracks in the tracking chambers and of
energy deposited in the electromagnetic and hadronic calorimeters of
the detector.
 
All tracking systems are located inside a solenoidal magnet which
provides a uniform axial magnetic field of 0.435~T along the beam
axis\footnote{In the OPAL coordinate system the $x$ axis points
  towards the centre of the LEP ring, the $y$ axis points upwards and
  the $z$ axis points in the direction of the electron beam.  The
  polar angle $\theta$ and the azimuthal angle $\phi$ are defined
  w.r.t. $z$ and $x$, respectively, while $r$ is the distance from the
  $z$-axis.}.
The magnet is surrounded by a lead glass electromagnetic
calorimeter and a hadron calorimeter of the sampling type.  Outside
the hadron calorimeter, the detector is surrounded by a system of muon
chambers.  There are similar layers of detectors in the forward and
backward endcaps.
 
The main tracking detector is the central jet chamber. This device is
approximately 4~m long and has an outer radius of about 1.85~m. It has 24
sectors with radial planes of 159 sense wires spaced by 1~cm. The
momenta $p$\ of
tracks in the $x$-$y$ plane are measured with a precision
$\sigma_p/p=\sqrt{0.02^2+(0.0015\cdot p[\mathrm{GeV}/c])^2}$.
 
The electromagnetic calorimeters in the barrel and the endcap sections
of the detector consist of 11704 lead glass blocks with a depth of
$24.6$ radiation lengths in the barrel and more than $22$ radiation lengths 
in the endcaps.
 
\section{ Data and Monte Carlo samples }
\label{sec_datamc}
The three data samples at $\rs\approx 172$, 183 and 
189 GeV that are used in this analysis were recorded as part
of the LEP-2 programme between 1996 and 1998. The luminosities,
evaluated using small angle Bhabha collisions, and the mean c.m.\ energies
are tabulated in Table~\ref{t:select}.

Monte Carlo event samples were generated at c.m.\ energies of 172.0,
183.0 and 189.0~GeV, including full simulation of the OPAL
detector~\cite{gopal}.  Events for the process
$\epem\rightarrow\zg\rightarrow\qqbar$, referred to as ``\zg\
events'', were generated using the PYTHIA 5.722~\cite{jetset3} parton
shower Monte Carlo code with initial and final state photon radiation and
fragmentation of the parton final state handled by the routines of
JETSET~7.408~\cite{jetset3}.  The simulation parameters have been
tuned to OPAL data taken at the \znull\ peak~\cite{OPALPR141}.
The JETSET Monte Carlo is able to provide a 
good description of experimental data for \epem\ annihilations with
c.m.\ energies from 10~GeV to
161~GeV~\cite{OPALPR158,alephas133,l3as133,delphias133,delphinch133,
bethke/pilcher}.

As an alternative fragmentation model, we generated
events with the HERWIG~5.9~\cite{herwig} parton shower Monte Carlo program,
also tuned to OPAL data, as described in~\cite{OPALPR197,OPALPR141}.
As an alternative model for initial state radiation (ISR)
the YFS3ff~3.6~\cite{yfsff} generator was used, coupled
to the same JETSET routines to handle the parton final states.

In addition we generated events of the type $\epem\rightarrow 4$
fermions (diagrams without intermediate gluons).  These 4-fermion
events, in particular those with four quarks in the final state,
constitute a major background for this analysis.  
Simulated 4-fermion events with quarks and leptonic final
states were generated using the GRC4F 1.2 Monte Carlo
Model~\cite{grc4f}.  The final states were produced via s-channel or
t-channel diagrams and include W$^+$W$^-$ events.
This generator is interfaced to JETSET~7.4 using
the same parameter-set for the parton shower, fragmentation and decays
as for \zg\ events.

Two other possible sources of background events were simulated.
Hadronic two-photon processes were
evaluated using PYTHIA, HERWIG and PHOJET~\cite{PHOJET}. Production of
$\epem\rightarrow\zg\rightarrow\tau\bar{\tau}$ was evaluated using
KORALZ~\cite{KORALZ}.

In addition to the Monte Carlo event generators discussed above, we
use the event generators ARIADNE 4.08~\cite{ariadne3} and
COJETS 6.23~\cite{cojets2}. ARIADNE is used to provide a
systematic check on the hadronization corrections for our \as\
measurement (Section~\ref{sec_alphas}). In contrast to the the other models,
COJETS is based on a parton shower model using independent
fragmentation and does not take coherence effects into account. 
Both COJETS and ARIADNE are used 
in addition to PYTHIA and HERWIG to compare to our corrected
data distributions. 
The parameter sets used for ARIADNE and COJETS are
documented in \cite{ariadne3} and \cite{cojetstuning}; both models
provide a good description of global \epem\ event properties at
$\rs=\mz$, as do PYTHIA and HERWIG.

\section{ Analysis method }
\label{sec_anal}
\subsection{ Selection of events }
Hadronic events are identified using criteria as described
in~\cite{OPALPR247}. The efficiency of selecting non-radiative hadronic events
is essentially unchanged with respect to lower c.m.\ energies and is
approximately 98\perc~\cite{OPALPR075}.  We define as particles
tracks recorded in the tracking chambers and clusters recorded in the
electromagnetic calorimeter.  The tracks are required to have
transverse momentum to the beam axis $p_T > 150$~MeV/$c$, a 
number of hits in the jet chamber $N_{hits} \geq 40$, a distance of
the point of closest approach to the collision point in the $r-\phi$
plane $d_0 \leq 2$ cm and in the $z$ direction $z_0 \leq 25$ cm.  The clusters in
the electromagnetic calorimeter are required to have a minimum energy
of 100~MeV in the barrel and 250~MeV in the endcap
sections~\cite{OPALPR158}.

To reject background from $\epem\rightarrow\tautau$ and
$\gamma\gamma\rightarrow\qqbar$ events and to ensure the events are
well contained in the OPAL detector we require at
least seven accepted tracks, and the cosine of the polar angle of the thrust axis
$|\costt|<0.9$.  The number of events after
this preselection, for each c.m.\ energy, are listed in
Table~\ref{t:select}.

To reject radiative events, we determine the effective c.m.\ energy \rsp\ of
the observed hadronic system as follows~\cite{OPALPR211}.  Isolated
photons in the electromagnetic calorimeter are identified, and the
remaining particles are formed into jets using the Durham~\cite{durham}
 algorithm with a value for the resolution parameter $\ycut=0.02$.
The energy of additional photons emitted close to the beam directions
is estimated by performing three separate kinematic fits assuming zero, one or two
such photons.  The probabilities of the fits are used to select the
most likely of these three.  The value of $\rsp$ is then
computed from the fitted momenta of the jets, excluding photons
identified in the detector or close to the beam directions.
The value of \rsp\ is set to \rs\ if the fit assuming zero  initial
state photons was selected. The 4-momenta of the measured particles
are then boosted into the rest frame of the observed hadronic system.
 
\begin{figure}[!htb]
\begin{center}
\resizebox{\textwidth}{!}
{\includegraphics{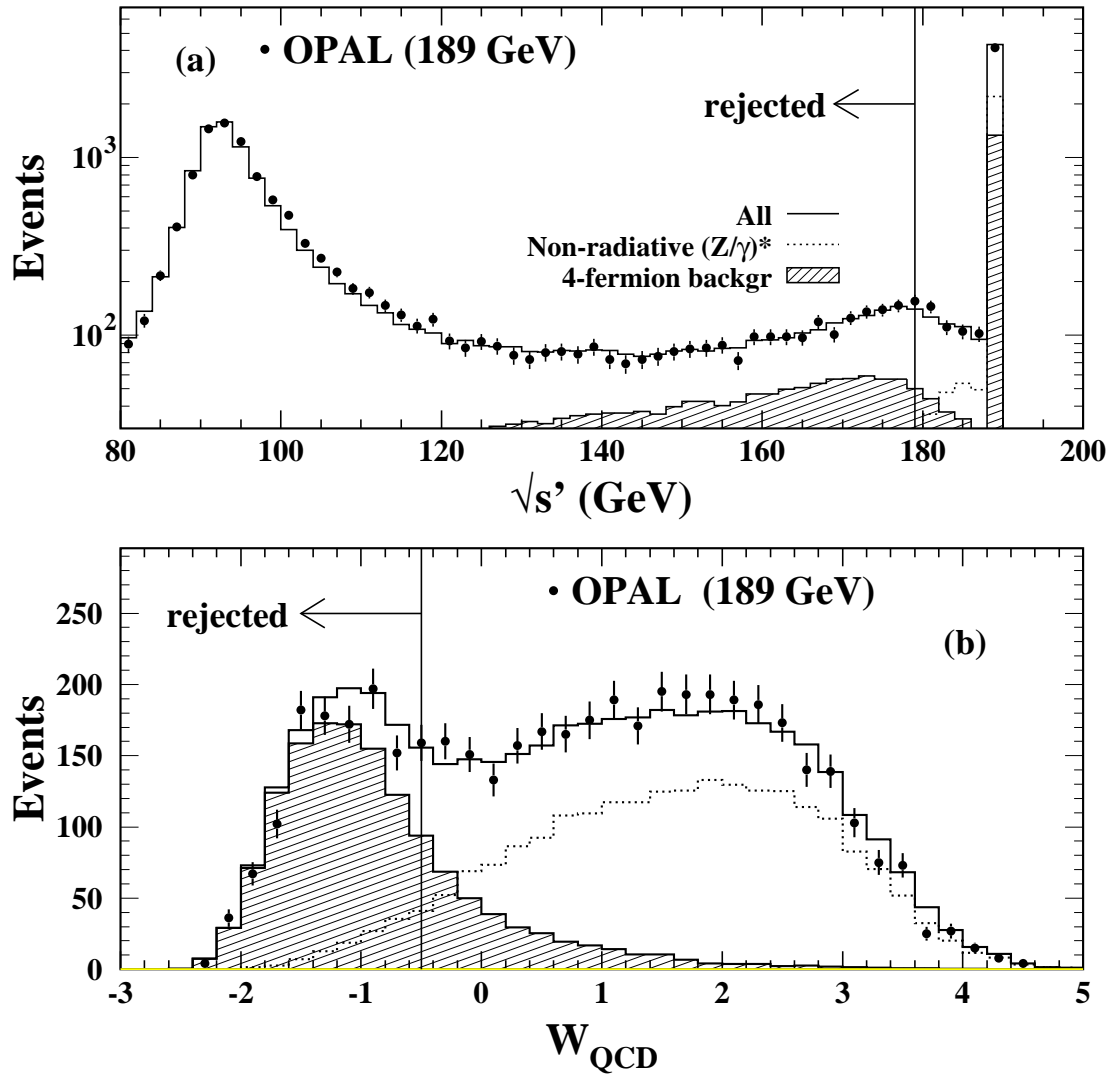}}
\caption[$s'$ distributions of signal and 4-fermion background]
{ a) Distribution of reconstructed \rsp\ for the data (full points) with
  statistical errors. The PYTHIA predictions for all \zg\ events
  (solid line) and for the non-radiative events,
  $\sqrt{s}-\sqrt{\sprime_{\mathrm{true}}}<1$~GeV, (dotted line) are
  also shown. The hatched area indicates the 4-fermion background
  predicted by the GRC4F Monte Carlo. 
  \\
  b) Distribution of the QCD event weight \wqcd. The notation is
  identical to that of a).  }
\label{fig_1}
\end{center}
\end{figure}

To reject events with large initial-state radiation (ISR), we require
$\rs-\rsp<10$~GeV.  This is referred to as the `ISR-fit' selection.
In Figure~\ref{fig_1}a we compare the $\rsp$ distribution of the data
to simulated \zg\ and 4-fermion events.  The simulated \zg\ events are
classified into radiative events with
$\sqrt{s}-\sqrt{\sprime_{\mathrm{true}}} > 1$~GeV (where
$\sqrt{\sprime_{\mathrm{true}}}$ is the true effective c.m.\ energy,
determined from Monte Carlo information), and non-radiative events,
which is the complement. While about $27\%$ of the selected 
\zg\ events are by this definition radiative events, the
fraction of events with $\sqrt{s}-\sqrt{\sprime_{\mathrm{true}}} > 10$~GeV
is only $5\%$.
The contributions of 4-fermion events,
simulated with GRC4F, are indicated separately.  The background from
4-fermion events and the efficiency of selecting non-radiative
events are given in Table~\ref{t:select}.  

The background from $\epem\rightarrow\tautau$ and
two-photon events of the type $\gamma\gamma\rightarrow\qqbar$ is
estimated from Monte Carlo samples to be less than 0.3\perc\ and is
neglected.

In order to reduce the background of 4-fermion events on the remaining
sample, we test the compatibility of the events with QCD-like
production processes. A QCD event weight \wqcd\ is computed 
as follows.  We force each event into a four-jet configuration in the
Durham jet scheme and use the EVENT2~\cite{event2} program to
calculate the ${\cal O}(\alpha_s^2)$ matrix element $\left| {\cal M}
(p_1,p_2,p_3,p_4) \right|^2 $ for the processes $\epem\rightarrow
\mathrm{q\bar{q}q\bar{q},q\bar{q}gg}$~\cite{ERT}. Since neither quark
nor gluon identification is performed on the jets, we calculate the
matrix element for each permutation of the jet-momenta and use the
permutation with the largest value for the matrix element to define
the event weight:
\begin{equation}
\wqcd\ = \max_{\left\{ p_1,p_2,p_3,p_4 \right\}} \log\left( \left| {\cal M}
(p_1,p_2,p_3,p_4) \right|^2 \right)
\end{equation}
with $p_i$ the momenta of reconstructed jets. Note that the 
definition of the event weight \wqcd\ contains kinematic information
only and is independent of the value of \as.

\begin{figure}[!htb]
\begin{center}
\resizebox{\textwidth}{!}
{\includegraphics{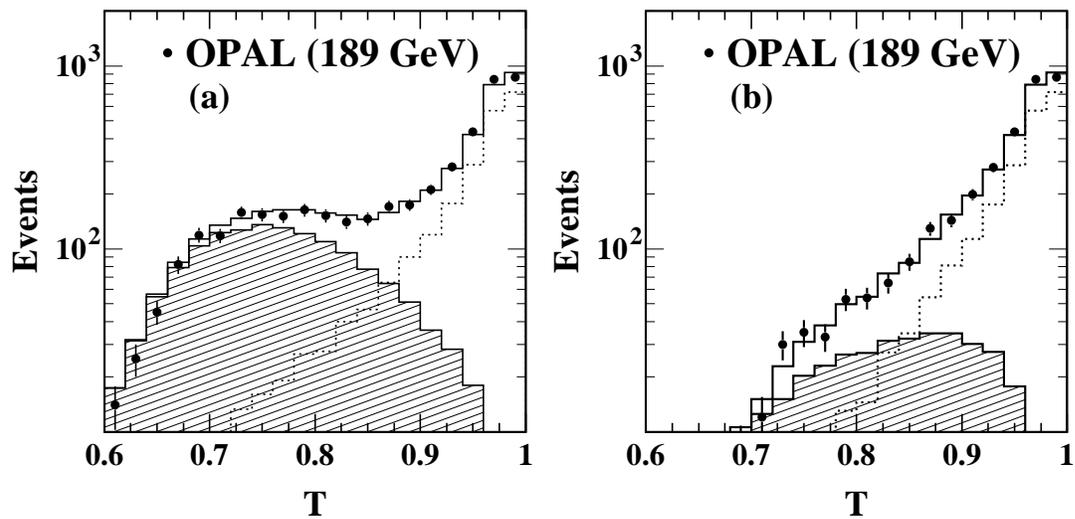}}
\caption[$s'$ distributions of signal and 4-fermion background]
{ The effect of the \wqcd\ selection cut on the thrust
  distribution (see Figure~\ref{fig_1} for
  notation). In (a) the thrust distribution, on detector level, is
  shown before the \wqcd\ selection is applied. In (b) the same, after
  the selection on \wqcd\ is applied. }
\label{fig_2}
\end{center}
\end{figure}

This weight \wqcd\ is expected to have large values for
processes described by the QCD matrix element, originating from
$\zg\rightarrow\qqbar$, and smaller values for \ww\ events. In
Figure~\ref{fig_1}b we compare the data distribution of \wqcd\ to the
expectations of our simulation.  A clear separation between the \zg\
and 4-fermion events is achieved by requiring $\wqcd \geq -0.5$.
This requirement reduces the 4-fermion background considerably, as can
be seen in Table~\ref{t:select}, whereas the efficiency for selecting
non-radiative \zg\ events is only slightly reduced.  In
Figure~\ref{fig_2} we show, as an example, the effect of the 4-fermion
background rejection on the distribution of thrust $T$, as defined in
Section~\ref{sec_shapes}.  The background from hadronic decays of \ww\ events
has predominantly low values for $T$, and is largely removed by our
selection.


\subsection{ Correction procedure }

The remaining 4-fermion background in each bin has been estimated by Monte Carlo
simulation and is subtracted from the observed bin content.  A bin-by-bin 
multiplication procedure is used to correct the
observed distributions for the effects of detector resolution and
acceptance as well as for the presence of remaining radiative \zg\
events.  For the multiplicity measurement presented in
Section~\ref{sec_mult}, we use a matrix correction procedure to
account for detector effects rather than a bin-by-bin procedure.

For the bin-by-bin correction procedure, each bin of each observable
is corrected from the ``detector level'' to the ``hadron level'' using
two samples of Monte Carlo \zg\ events at each c.m.\ energy.  The
hadron level does not include initial state radiation or detector
effects and allows all particles with lifetimes shorter than $3\times
10^{-10}$~s to decay.  The detector level includes full simulation of
the OPAL detector and initial state radiation and contains only those
events which pass the same cuts as are applied to the data.  The
bin-by-bin correction factors are derived from the ratio of the
distributions at the hadron level to those at the detector level.

A bin-by-bin correction procedure is suitable for most quantities as
the effects of finite resolution and acceptance do not cause
significant migration (and therefore correlation) between bins.  For
the multiplicity measurement however, such a method is not readily applicable,
due to the large correlations between bins.


The four-momenta of tracks
and of the electromagnetic calorimeter clusters not associated with
tracks were used to calculate event shapes.  When a calorimeter cluster had associated
tracks, their expected energy deposition was used to reduce
double counting by correcting the cluster energy.  If the energy of a
cluster was smaller than the expected energy response of the
associated tracks, the cluster energy was not used. The masses of all
charged particles were set to the pion mass and the invariant masses
of the energy clusters were assumed to be zero.

%
%
%
\subsection{ Systematic uncertainties }
\label{sec_syserr}

The experimental systematic uncertainty is estimated by repeating the
analysis with varied experimental conditions.
In order to reduce bin-to-bin fluctuations in the magnitudes of the
systematic uncertainties we average the relative uncertainty over
three neighbouring bins.

To allow for any inconsitencies caused by differences between the responses of the 
tracking or the calorimeter, three differences are formed for the quantities in each 
bin; between the standard result, the one obtained using tracks and all clusters 
and the one obtained using only tracks. The largest of them is taken as the 
systematic error.

The inhomogeneity of the response of the detector in the endcap region
was allowed for by restricting the analysis to the barrel region of the
detector, requiring the thrust axis of accepted events to lie within
the range $|\costt|<0.7$. The event sample is reduced in size by
approximately 27\perc.  The corresponding systematic error is the deviation of the
results from those of the standard analysis.

For observables measured using information from charged particles
only, we evaluate an additional uncertainty on the track modelling.
The maximum allowed distance
of the point of closest approach of a track to the collision point in
the $r$-$\phi$ plane $d_0$ was changed from 2 to 5~cm,
the maximal distance in the 
$z$ direction $z_0$ from 25 to 10 cm and the minimal number of hits
from 40 to 80. These modifications change the number of tracks
by up to approximately 12\perc. The quadratic sum of the deviations from 
the standard result is taken to be the systematic error due to the 
uncertainty of the track modelling.

Uncertainties arising from the selection of non-radiative events are
estimated by repeating the analysis using a different
technique\cite{OPALPR197} to determine the value for \rsp.  This
technique differs from our standard \rsp\ algorithm in that in this case
the kinematic fit assumes always one unobserved photon close to the beam
direction for each event.  The final event sample with this \rsp\
algorithm has an overlap of approximately 94\perc\ with the standard
sample, and is reduced in size by 3\perc.  The difference relative to
the standard result is taken as the systematic error.

We evaluated our \zg\ selection efficiencies using
events generated with YFS3ff, which contains QED exponentiated matrix
elements for ISR up to {\cal O}($\alpha_{EM}^3$). Similar estimates of
efficiencies are obtained if this model is used instead of PYTHIA, and
therefore no additional systematic uncertainty is assigned.

Systematic uncertainties associated with the subtraction of the
4-fermion background events are estimated by varying the position
of the cut-value on the QCD event weight \wqcd\ (see Figure~\ref{fig_1}b).
We vary this position to $\wqcd\geq-0.8$ which increases the event sample by
approximately 7\perc, and $\wqcd\geq 0$ which decreases the event
sample by approximately 12\perc. We take the maximum deviation from
our standard result $\wqcd\geq-0.5$ as the systematic uncertainty.

In addition we vary the predicted background to be subtracted,
within its measured uncertainty of 5\perc~\cite{OPALPR260}, and use
the largest difference from the standard result as a systematic uncertainty.

The difference in the results when we use simulated \zg\ events
generated using HERWIG instead of PYTHIA is taken as the uncertainty
in the modelling of the \zg\ events.


\section{ Results }
\label{sec_results}
\subsection{ Event shapes }
\label{sec_shapes}
The properties of hadronic events may be characterised by a set of
event shape observables.  The following quantities are considered:
\begin{description}
\item[Thrust \bm{T}:]
  defined by the expression\cite{thrust1,thrust2}
  \begin{equation}
  T= \max_{\vec{n}}\left(\frac{\sum_i|p_i\cdot\vec{n}|}
                    {\sum_i|p_i|}\right)\;\;\;.
  \label{equ_thrust}
  \end{equation}
  The thrust axis $\vec{n}_T$ is the direction $\vec{n}$ which
  maximises the expression in parenthesis.  A plane through the origin
  and perpendicular to $\vec{n}_T$ divides the event into two
  hemispheres $H_1$ and $H_2$.
\item[Thrust major \bm{\tma}.]
The maximisation in equation~\ref{equ_thrust} is performed with  
 the condition that
  $\vec{n}$ must lie in the plane perpendicular to $\vec{n}_T$. The
  resulting vector is called $\vec{n}_{\tma}$.
\item[Thrust minor \bm{\tmi}.]
  The expression in parenthesis  is evaluated for the vector
  $\vec{n}_{\tmi}$ which is perpendicular both to $\vec{n}_T$ and to
  $\vec{n}_{\tma}$.
\item[Oblateness \bm{O}.] 
  This observable is defined by $O=\tma-\tmi$ \cite{def_o}.
\item[Sphericity \bm{S} and Aplanarity \bm{A}.] 
  These observables are based on the momentum tensor
  \[
    S^{\alpha\beta}= \frac{\sum_ip_i^{\alpha}p_i^{\beta}}{\sum_ip_i^2}\;\;\;,
    \;\;\;\alpha,\beta= 1,2,3\;\;\;.
  \]
  The three eigenvalues $Q_j$ of $S^{\alpha\beta}$ are ordered such
  that $Q_1<Q_2<Q_3$. These then define $S$ \cite{def_s1,def_s2} and $A$
  \cite{def_a} by
  \[
    S= \frac{3}{2}(Q_1+Q_2)\;\;\;\mathrm{and}\;\;\; A= \frac{3}{2}Q_1\;\;\;.
  \]
\item[C-parameter.]
  The momentum tensor $S^{\alpha\beta}$ is linearised to become
  \[
    \Theta^{\alpha\beta}= \frac{\sum_i(p_i^{\alpha}p_i^{\beta})/|p_i|}
                               {\sum_i|p_i|}\;\;\;,
                           \;\;\;\alpha,\beta= 1,2,3\;\;\;.
  \]
  The three eigenvalues $\lambda_j$ of this tensor define \cp
  \cite{def_c} with
  \[
    \cp= 3(\lambda_1\lambda_2+\lambda_2\lambda_3+\lambda_3\lambda_1)\;\;\;.
  \]
\item[Heavy Jet Mass \bm{\mh}.] The hemisphere 
invariant masses are calculated using  the particles
  in the two hemispheres $H_1$ and $H_2$.   We define
  \mh\ \cite{def_mh1,def_mh2} as the heavier mass, divided by $\rs$ .
\item[Jet Broadening variables \bm{\bt} and \bm{\bw}.] 
  These are defined by computing the quantity
  \[
    B_k= \left(\frac{\sum_{i\in H_k}|p_i\times\vec{n}_T|}
                    {2\sum_i|p_i|}\right)
  \label{equ_btbw}
  \]
  for each of the two event hemispheres, $H_k$,  defined above.
  The two observables \cite{nllabtbw} are defined by
  \[
    \bt= B_1+B_2\;\;\;\mathrm{and}\;\;\;\bw= \max(B_1,B_2)\;\;\;
  \]
  where \bt\ is the total and \bw\ is the wide jet broadening.
\item[Transition value between 2 and 3 jets {\boldmath \ytwothree}.]
  The value of \ycut\ for the Durham jet scheme at which the
  event makes a transition between a  2-jet and a 3-jet assignment.
\end{description}
 
In the following, we use the symbol $y$ to denote a generic event
shape observable, where larger values of $y$ indicate regions
dominated by the radiation of hard gluons and small values of $y$
indicate the region influenced by multiple soft gluon radiation. 
Note that thrust $T$ forms an exception to this rule, as the value
of $T$ reaches one for events consisting of two collimated,
back-to-back, jets. We therefore occasionally use \thr\ instead. 

\begin{figure}[!htb]
\resizebox{\textwidth}{!}
{\includegraphics{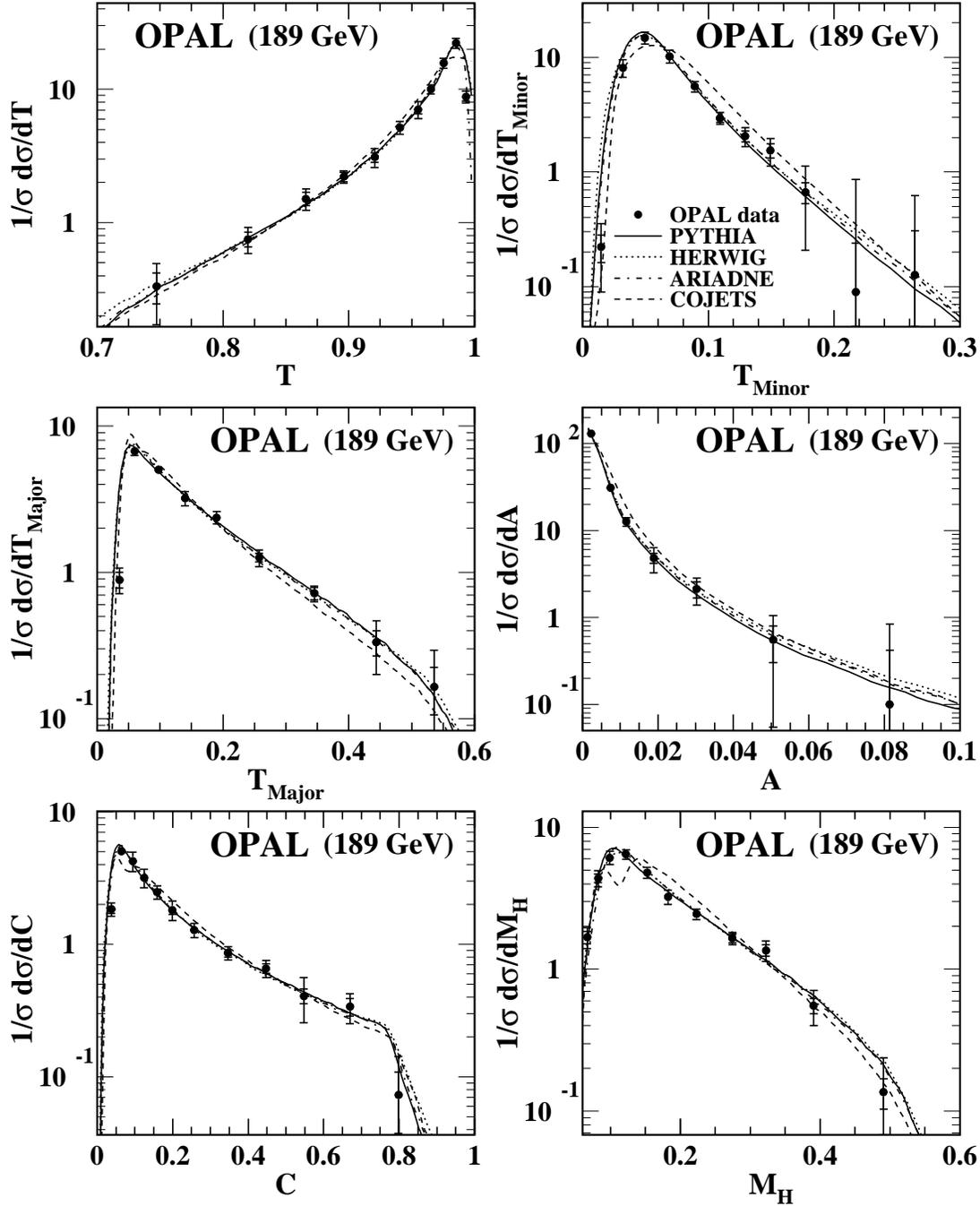}}
\caption[]
{ Distributions at $\rs=189$~GeV of the event shape observables thrust
  $T$, thrust major \tma, thrust minor \tmi, aplanarity $A$,
  C-parameter \cp\ and heavy jet mass \mh. Experimental statistical
  errors are delimited by the inner small horizontal bars. The total
  errors are shown by the outer horizontal error bars.  Hadron level
  predictions from PYTHIA, HERWIG and ARIADNE are also shown.}
\label{f:evsh1}
\end{figure}

\begin{figure}[!htb]
\resizebox{\textwidth}{!}
{\includegraphics{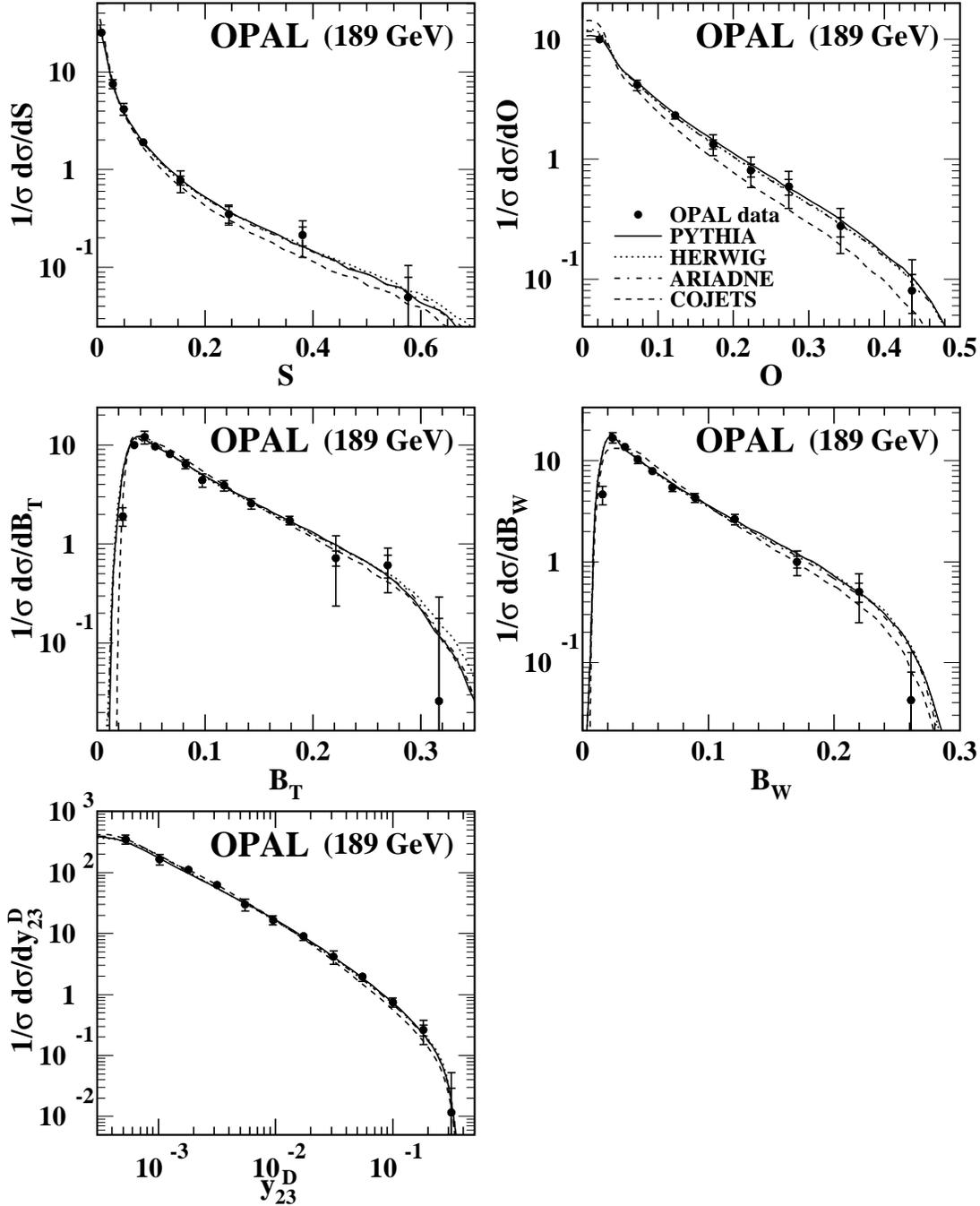}}
\caption[]
{ Distributions at $\rs=189$~GeV of the event shape observables
  sphericity $S$, oblateness $O$, total jet broadening \bt, wide jet
  broadening \bw, and the transition value between 2- and
  3-jets \ytwothree.
  Experimental statistical errors are delimited by the
  inner small horizontal bars. The total errors are shown by the outer
  error bars.  Hadron level predictions from PYTHIA, HERWIG
  and ARIADNE are also shown.}
\label{f:evsh2}
\end{figure}

Figures~\ref{f:evsh1} and~\ref{f:evsh2} show the distributions of the
event shape observables $T$, \tma, \tmi, $A$, \cp, \mh, $S$, $O$,
\bt, \bw, and \ytwothree, corrected for detector acceptance and initial
state radiation, plotted at the weighted centres
of the bins.  The data obtained at 189~GeV  are shown with the 
statistical  and systematic
uncertainties added in quadrature.  The numerical values for all the
event shape distributions, for c.m.\ energies of 172, 183 and 189~GeV,
are listed in Tables~\ref{t:evsset1} to \ref{t:evsset3}. 

With the statistics available at  189~GeV, all event shapes
are determined with high precision. There are no deviations
from the predictions of PYTHIA, ARIADNE and HERWIG.
The COJETS Monte Carlo model deviates from the data for \tmi,
\mh\ and $O$, but describes the remaining event shapes reasonably well.
Note that all generators have been tuned at $\rs=\mz$, and are not
re-tuned at these higher energies.

\begin{figure}[!htb]
\resizebox{\textwidth}{!}
{\includegraphics{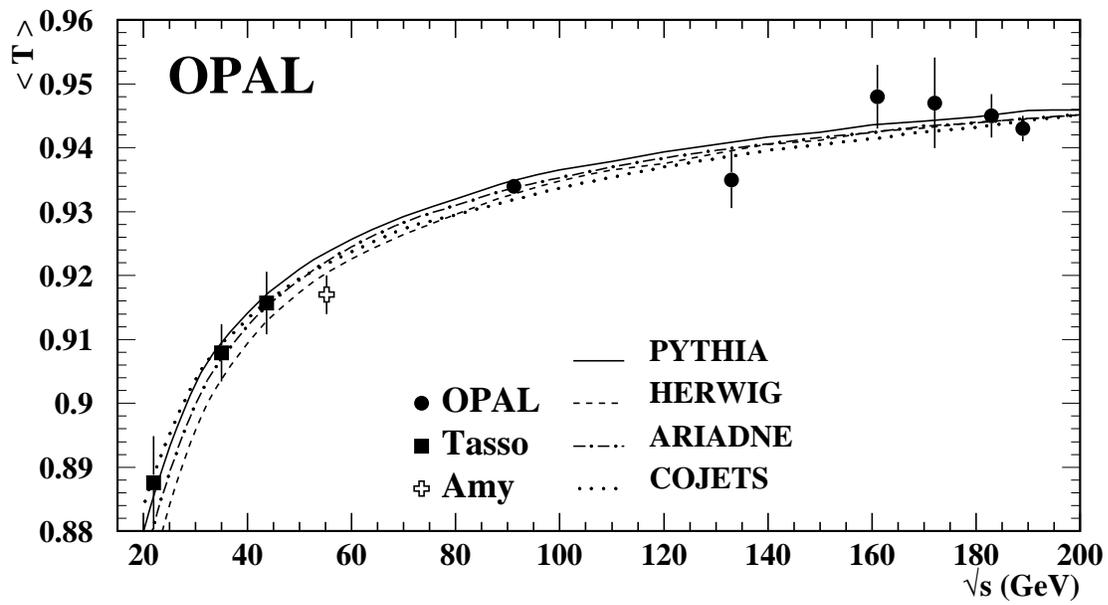}}
\caption[]
{ Distribution of the mean thrust $\left< T \right>$ as a
  function of the centre-of-mass energy, compared to several
  QCD models.
} 
\label{f:thrust}
\end{figure}

The mean value of the thrust distribution, $\langle T \rangle$, as a
function of the c.m. energy, is shown in Figure~\ref{f:thrust}
together with data from lower energy
measurements~\cite{OPALPR158,OPALPR197,OPALPR075,thrust_data}
and the predictions for the energy evolution from PYTHIA, HERWIG,
ARIADNE and COJETS. The predictions of all four Monte Carlo models are
consistent with our new measurements.

%
%
%
\subsection{ Determination of \bm{\as} }
\label{sec_alphas}
Our measurement of the strong coupling strength \as(Q) is based on
fits of the QCD predictions to the corrected distributions for \thr,
\mh, \cp, \bt, \bw\ and \ytwothree.
The theoretical descriptions of
these six observables are the most complete, allowing the use of
combined \oaa+NLLA QCD
calculations~\cite{nllathmh,nllabtbw,nllad2,nllabtbw2,nllacp}.
We follow the procedures described in
references~\cite{OPALPR075,OPALPR158} as closely as possible in order
to obtain results which we can compare directly to our previous
analysis.  In particular, we choose the so-called \lnr-matching
scheme, and fix the renormalization scale parameter, $\xmu\equiv
\mu/\rs$, to $\xmu=1$, where $\mu$ is the energy scale at which the
theory is renormalised.  The \oaa+NLLA prediction for
\cp\ has only recently become available~\cite{nllacp}, and was not
used in our earlier publications.

Analytic QCD predictions describe distributions at the level of quarks
and gluons (parton level).  The predictions are convolved with
hadronization effects by multiplying by the ratio of hadron- to
parton level distributions determined using a Monte Carlo
model. We use PYTHIA to generate events at $\rs=172$, 183 and
189~GeV for this purpose.

The fit ranges are determined by the following considerations. For
each observable, the ratio between Monte Carlo distributions computed
for partons and hadrons is required to be unity to within about
10\perc\ and the distribution of
partons from Monte Carlo models is required to be  described well by
the analytic predictions. The fit ranges are identical to the fit ranges
used in our studies at $\rs=\mz$, 130 and
161~GeV~\cite{OPALPR075,OPALPR158,OPALPR197}.

\begin{figure}[!htb]
\resizebox{\textwidth}{!}
{\includegraphics{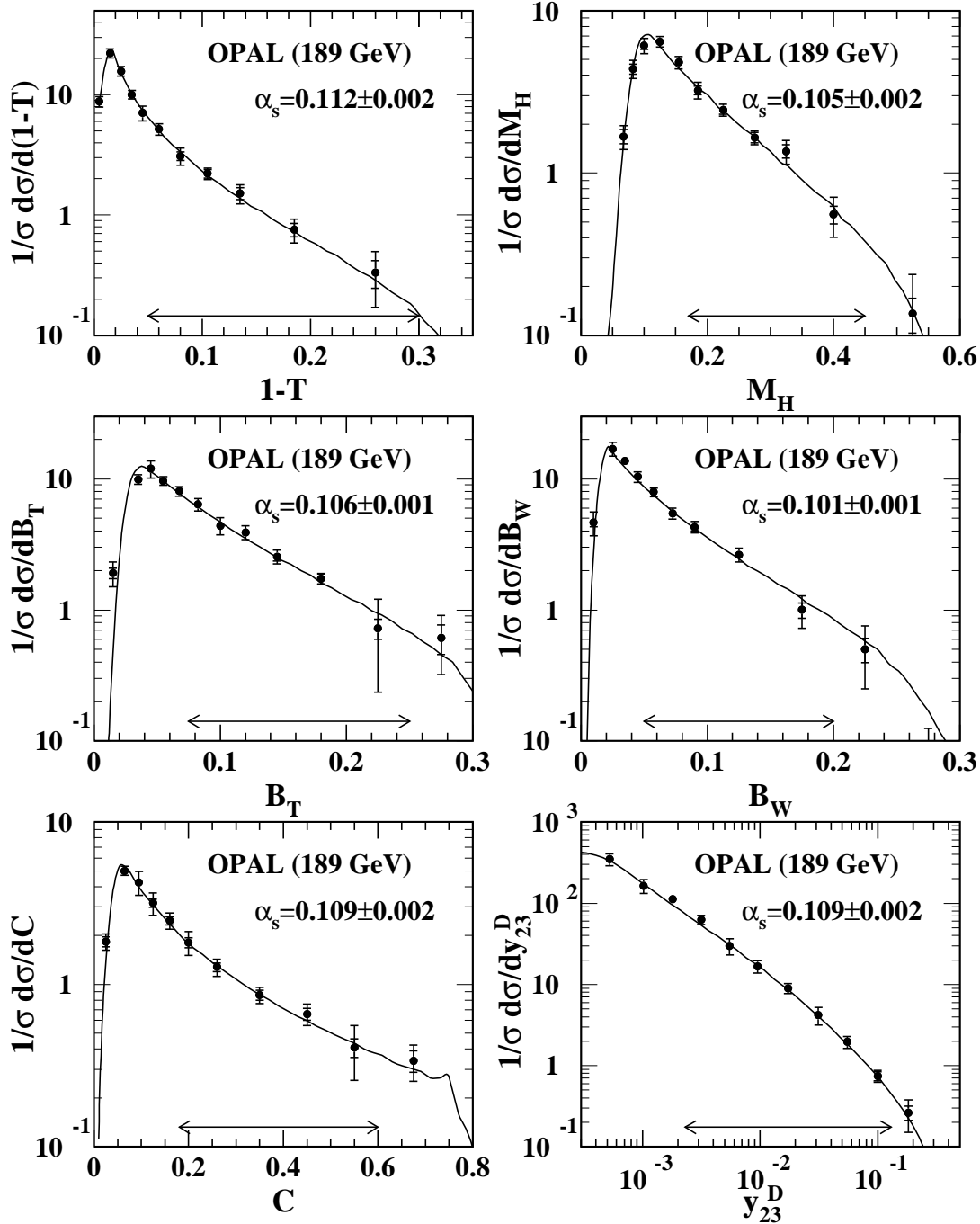}}
\caption[]  
{ Distributions of the event shape observables thrust \thr, heavy jet
  mass \mh, total \bt\ and wide \bw\ jet broadening, \cp-parameter and
  the transition value between 2- and 3-jets \ytwothree, are shown
  together with fits of the \oaa+NLLA QCD predictions, with \as\ as fitted parameter.  
  The fitted regions are indicated by the arrows.  }
\label{fig_asresults}
\end{figure}

We find satisfactory fits for all six observables, for all three c.m.\
energies. In particular, the \cp\ distributions are well described by the
newly introduced predictions.  In Table~\ref{tab_asresults} we quote
the result on \as(172~GeV), \as(183~GeV) and \as(189~GeV) for all six
observables.  In Figure~\ref{fig_asresults} we show the results for
the determination of \as(189~GeV).  Note also that outside the fit
ranges the data agree with the \oaa+NLLA predictions.

For each observable, the statistical uncertainties are estimated from
the variance of \as\ values derived from fits to 100 independent sets
of simulated events, each with the same number of events as the data.

We also derive a combined result of the six observables for the strong
coupling strength using the weighted average procedure as described in
reference~\cite{OPALPR075}, which includes the correlations between
the shapes.  The statistical uncertainty of the combined result is
estimated using 100 independent samples of Monte Carlo events in the
same manner as for the individual measurements.

The experimental uncertainty is estimated by adding in quadrature the
following contributions: the largest difference between the central
result and the results from tracks and all clusters or tracks only; the
difference found when using the alternative \rsp\ selection; the
difference when requiring the thrust axis to lie in the range
$|\cos\theta_T|<0.7$; the difference from using the variation on
\wqcd\ to reject 4-fermion events; and the difference when the
4-fermion background is scaled by $\pm$5\perc.
The experimental uncertainties, shown for each observable in
Table~\ref{tab_asresults}, are of similar size as the statistical
uncertainties. In Table~\ref{tab_asmean} we present details of the
experimental uncertainty for the weighted mean value of \as, at each
c.m.\ energy.

The hadronization uncertainty is defined by adding in quadrature: the
larger of the changes in \as\ observed when varying the hadronization
parameters $b$ and $\sigma_Q$ by $\pm$~1 standard deviation about
their tuned values~\cite{OPALPR141} in PYTHIA; the change observed
when the parton virtuality cutoff parameter is altered from
$Q_0=1.9$~GeV to $Q_0=4$~GeV in PYTHIA (without changing the other
Monte Carlo parameters), corresponding to a change of the mean parton
multiplicity from 7.4 to 5.0; the change observed when at the parton
level only the light quarks u, d, s and c are considered in order to
estimate potential quark mass effects; and both differences with
respect to the standard result when HERWIG or ARIADNE are used to
account for the hadronization effects, rather than PYTHIA.  For all
observables the hadronization uncertainties are given in
Table~\ref{tab_asresults}.  These uncertainties are relatively small
compared to the statistical uncertainty of \as\ for the 172~GeV
sample, but become of similar size to the statistical uncertainty for
the 189~GeV sample. In Table~\ref{tab_asmean} we show the individual
contributions to the hadronization uncertainty for the weighted mean
value of \as.

 
The importance of uncomputed higher order terms in the theory may be
estimated by studying the effects of varying the renormalization scale
parameter \xmu. We estimate the dependence of our fit results on the
renormalization scale \xmu\ as in
references~\cite{OPALPR075,OPALPR158}, by repeating the fits using
$\xmu=0.5$ and $\xmu=2$. We define the average deviation from the central 
result as the systematic uncertainty.
 We find variations which are generally
larger than any other systematic variation and which are highly
correlated between all observables\footnote{ 
The compensation in \as\ due to a change of the renormalisation scale
\xmu\ in the QCD predictions is proportional to the value of \as\ itself. 
The scale uncertainty at our 172~GeV sample is therefore smaller than that
at 183 or 189~GeV, where the value for \as\ we obtain is larger. }.
   
The total uncertainty for each individual observable is computed by adding
in quadrature the statistical, experimental, hadronization and
scale uncertainties. The total uncertainty on \as\ is typically
around 10\perc\ at $\rs=172$ GeV and around 5\perc\ at $\rs=183$ and $\rs=189$ GeV.  

As a consistency check we repeated the fits to \as\ while varying the
fit ranges. The fit range was changed by plus or minus one bin at
either side. In general the observed deviations to the central value
of \as\ were small and we do not include this
check in our systematic uncertainty.

\begin{figure}[!htb]
\resizebox{\textwidth}{!}
{\includegraphics{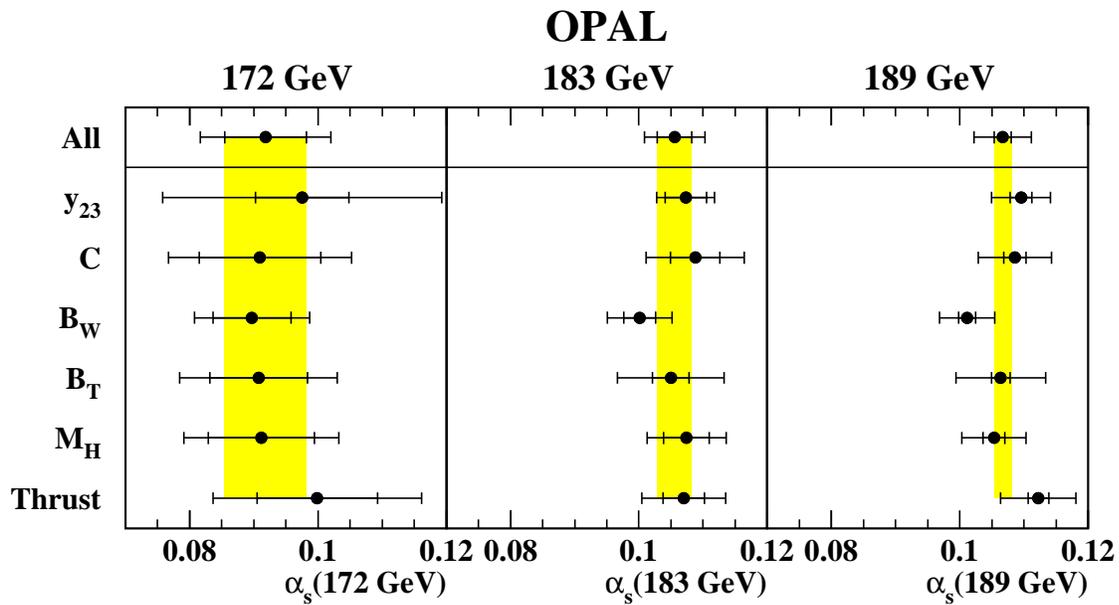}}
\caption[]  
{ Fitted values of \as\ for each event shape at 172, 183 and 189~GeV.
  The inner error bar is statistical, the outer corresponds to the total 
  systematic uncertainty. The light band correspons to the statistical
  uncertainty of \as\ obtained as the weighted mean.  
}  
\label{fig_asresum}
\end{figure}

Our result for the computed average of the six event shapes, each weighted with
its total uncertainty, is
\begin{eqnarray}
 \label{e_alphas}
  \as(172~\gev) &=& 0.092\pm0.006\stat\pm0.008\syst , \nonumber \\
  \as(183~\gev) &=& 0.106\pm0.003\stat\pm0.004\syst , \\
  \as(189~\gev) &=& 0.107\pm0.001\stat\pm0.004\syst . \nonumber 
\end{eqnarray}
The systematic uncertainty has contributions from experimental effects
$\pm(0.002-0.007)$, hadronization effects $\pm(0.001-0.002)$ and from
variations of the renormalization scale $\pm(0.002-0.004)$, as
explained above.  The systematic variations of the combined results
are detailed in Table~\ref{tab_asmean}.  Their values evolved to \mz\
are listed in Table~\ref{tab_asmz}. In Figure~\ref{fig_asresum}
we show the values of \as\ we obtain from all event shapes at the
three energies. In the same figure we also show the computed weighted
averages of \as.  The spread in the values of \as\ as obtained from
the six event shapes is small, and the computed weighted average of
\as\ covers the individual measurements within the
uncertainty. Note however that \as\ determined using \bw\ at 183 and
189 GeV is somewhat smaller, as already observed at LEP-1 energies~\cite{OPALPR075}, 
which may indicate that the higher order
corrections to \bw\ are slightly different from those of the other event
shapes.

The correlations
between the values of \as\ at the three different c.m.\ energies are large.
For example, the correlations
between the theoretical uncertainties of the three values of \as\
quoted in equation~\ref{e_alphas}, are all close to 100\perc. The
correlation coefficients for the experimental uncertainties vary
between 40\perc\ and 67\perc, and the ones for the hadronization
corrections vary between 84\perc\ and 98\perc.

In order to allow for the effects of these correlations, we construct
a value of \as\ at the scale given by the luminosity weighted c.m.\
energy of our three data samples, 186.7 GeV.  We use \oaaa\ predictions
to evolve all our values of \as\ to this scale. As a final result,
using the weighted mean values of \as\ at all three energies, we
obtain
\begin{eqnarray}
 \label{e_alphasfin}
 \as(187~\gev) &=& 0.106\pm0.001\stat\pm0.004\syst .
\end{eqnarray}

%
%
When evolved to the scale of \mz, using \oaaa\ calculations, the
value of the weighted average for \as\ becomes
$\asmz=0.117\pm0.005$ (see also Table~\ref{tab_asmz}).
For comparison, our measurements at the \mz\
energy with a slightly different set of observables based on \oaa+NLLA
QCD calculations yielded $\asmz=0.120\pm0.006$~\cite{OPALPR075}.

In order to compare average \as\ values using exactly the same
theoretical predictions and observables as used in our previous publications, 
we compute the weighted average from our present fits to only
\thr, \mh, \bt\ and \bw. This is 
$\as(187~\gev)=0.104\pm0.005$ which corresponds to
$\asmz=0.115\pm0.006$ when evolved to the scale of \mz.  Our previous
analysis at $\rs=\mz$ gave $\asmz=0.116\pm0.006$ from fits of the same
predictions to the same observables.  Our present determinations of
\as\ are therefore consistent with our measurement at $\rs=\mz$.  

\begin{figure}[!htb]
\resizebox{\textwidth}{!}
{\includegraphics{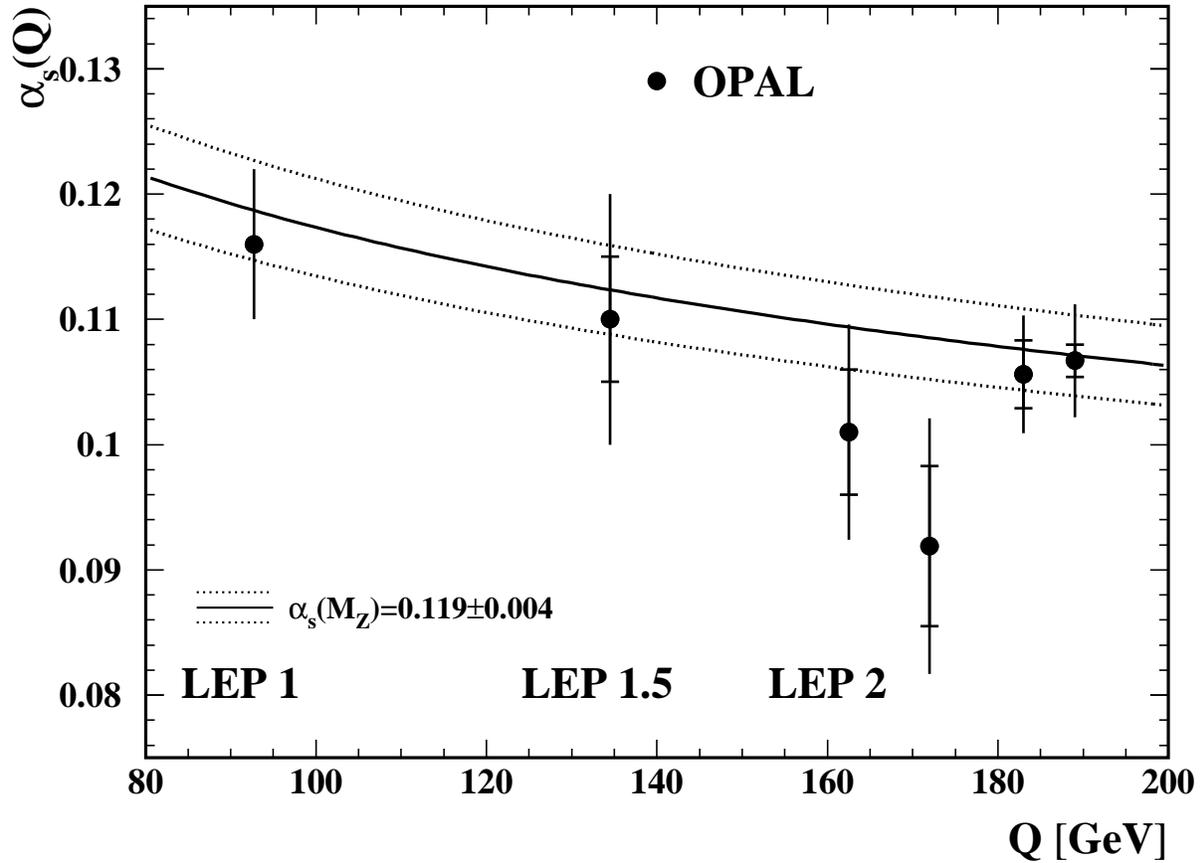}}
\caption[]
{ Values of \as, as determined from fits to \thr, \mh, \bt\ and \bw,
  as function of energy, on a linear scale.  The curves show the
  \oaaa\ QCD prediction for \asq\, using
  $\asmz=0.119\pm0.004$~\protect{\cite{bethke98}}; the full line shows
  the central value while the dotted lines indicate the variation
  given by the uncertainty. }
\label{fig_asrun}
\end{figure}

In Figure~\ref{fig_asrun} we show our result~(\ref{e_alphas}) together
with our other measurements of the strong coupling strength, as a
function of the energy scale $Q$.  The curve shows the \oaaa\
prediction for \asq\, using $\asmz=0.119\pm0.004$, the value of \asmz\
as given in reference~\cite{bethke98}.  Note that this value of \as\
is obtained using many observables, from a wide variety of
experimental environments. Our determinations of \as, obtained with
the four event shapes mentioned above, all fall below the predictions
of reference~\cite{bethke98}. This indicates that the values obtained
using these observables are somewhat shifted with respect to the value
from reference~\cite{bethke98}.  However, the consistency of our
measurements between the LEP-1 and LEP-2 c.m.\ energies is excellent.
 
We obtained a relatively low value of \as\ using the 172~GeV sample alone
for all event shape observables; approximately 1.8 standard
deviations below expectations. 

\begin{figure}[!htb]
\resizebox{\textwidth}{!}
{\includegraphics{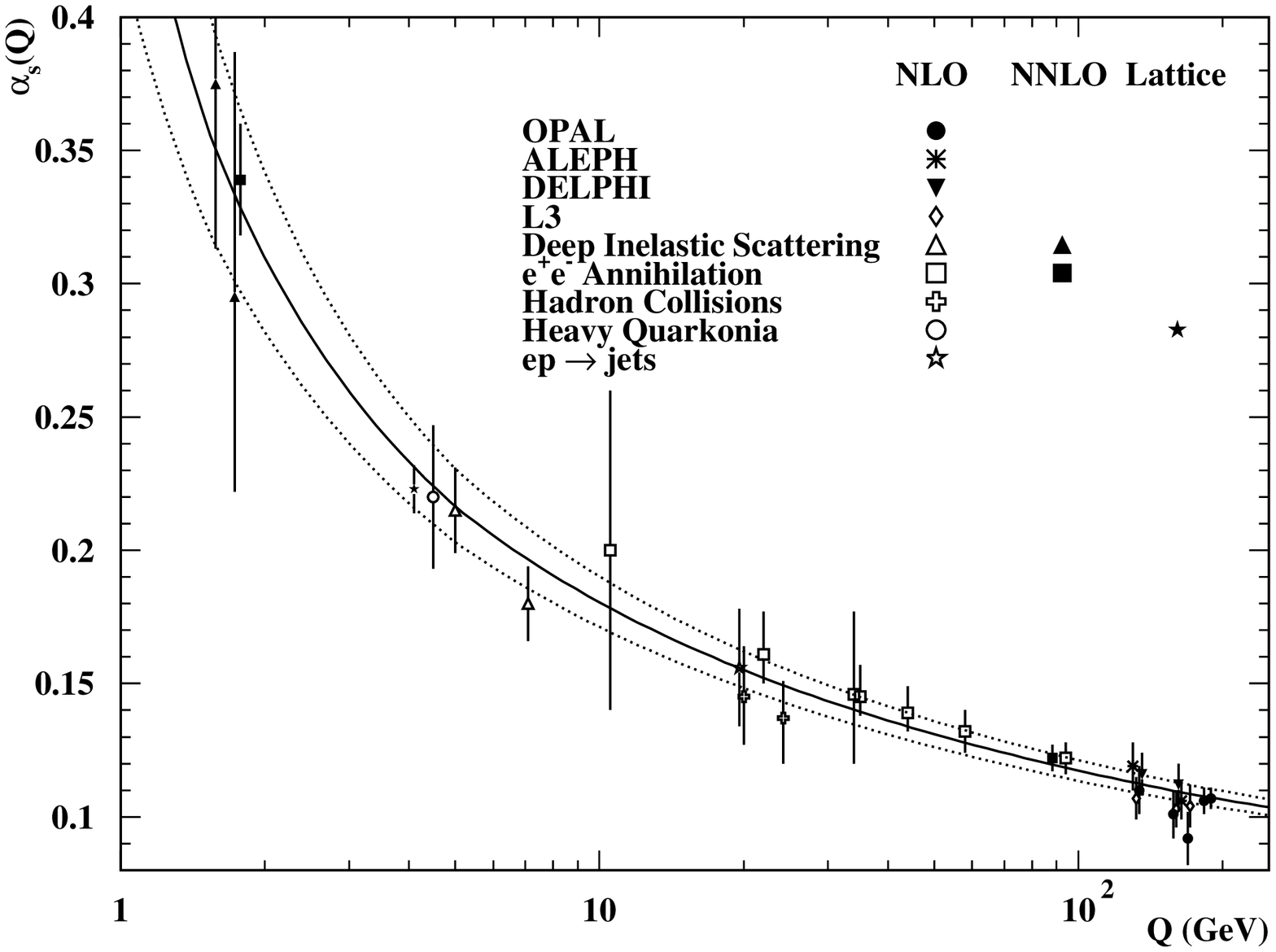}}
\caption[]
{ Values of \as\ as function of energy~\protect{\cite{bethke98}}. The
  labels NLO and NNLO refer to the order of calculation used. NLO
  corresponds to \oaa\ in \epem\ annihilations, and NNLO to \oaaa. The
  label Lattice refers to \as\ values determined from lattice QCD
  calculations. The curves show the \oaaa\ QCD prediction for \asq\,
  using $\asmz=0.119\pm0.004$; the full line shows the central value
  while the dotted lines indicate the variation given by the
  uncertainty. }
\label{fig_aslog}
\end{figure}

In Figure~\ref{fig_aslog} we show our results again, on a logarithmic 
scale, together with results obtained from the other experiments 
at lower c.m.\ energies. 

\subsection{ Charged Multiplicity }
\label{sec_mult}
We measure the charged particle multiplicity distribution and
derive several related quantities from it, in particular
the mean charged multiplicity \mnch, the dispersion $D
=(\langle n_{\mathrm{ch}}^2\rangle- 
\langle n_{\mathrm{ch}}\rangle^2)^{1\over 2}$, the ratio 
$\langle n_{\mathrm{ch}}\rangle/D$, 
the normalised second moment 
$C_{2}
=\langle n_{\mathrm{ch}}^{2}\rangle/\langle n_{\mathrm{ch}}\rangle^{2}$ 
and the second binomial moment~\cite{OPALPR197} 
$\rtwo
=\langle n_{\mathrm{ch}}(n_{\mathrm{ch}}-1)\rangle/\mnch^2$.

To correct the observed charged particle multiplicity distribution, 
we first subtract the expected background from 4-fermion events as described
in Section~\ref{sec_anal}. The resulting distribution is corrected
for experimental effects such as acceptance, resolution and secondary
interactions in the detector with an unfolding matrix, as previously done 
in references~\cite{OPALPR158,OPALPR046}. This matrix is
determined from the PYTHIA sample of fully simulated \zg\ events.
Biases introduced by the event selection, by radiative events passing
the selection and by the fraction of 
particles with lifetimes  shorter than $3\times
10^{-10}$~s that did not decay in the detector
 are corrected using a bin-by-bin multiplication method. The
bin-by-bin corrections are typically smaller than 10--15\perc.
 
\begin{figure}[!htb]
\resizebox{19cm}{!}
{\includegraphics{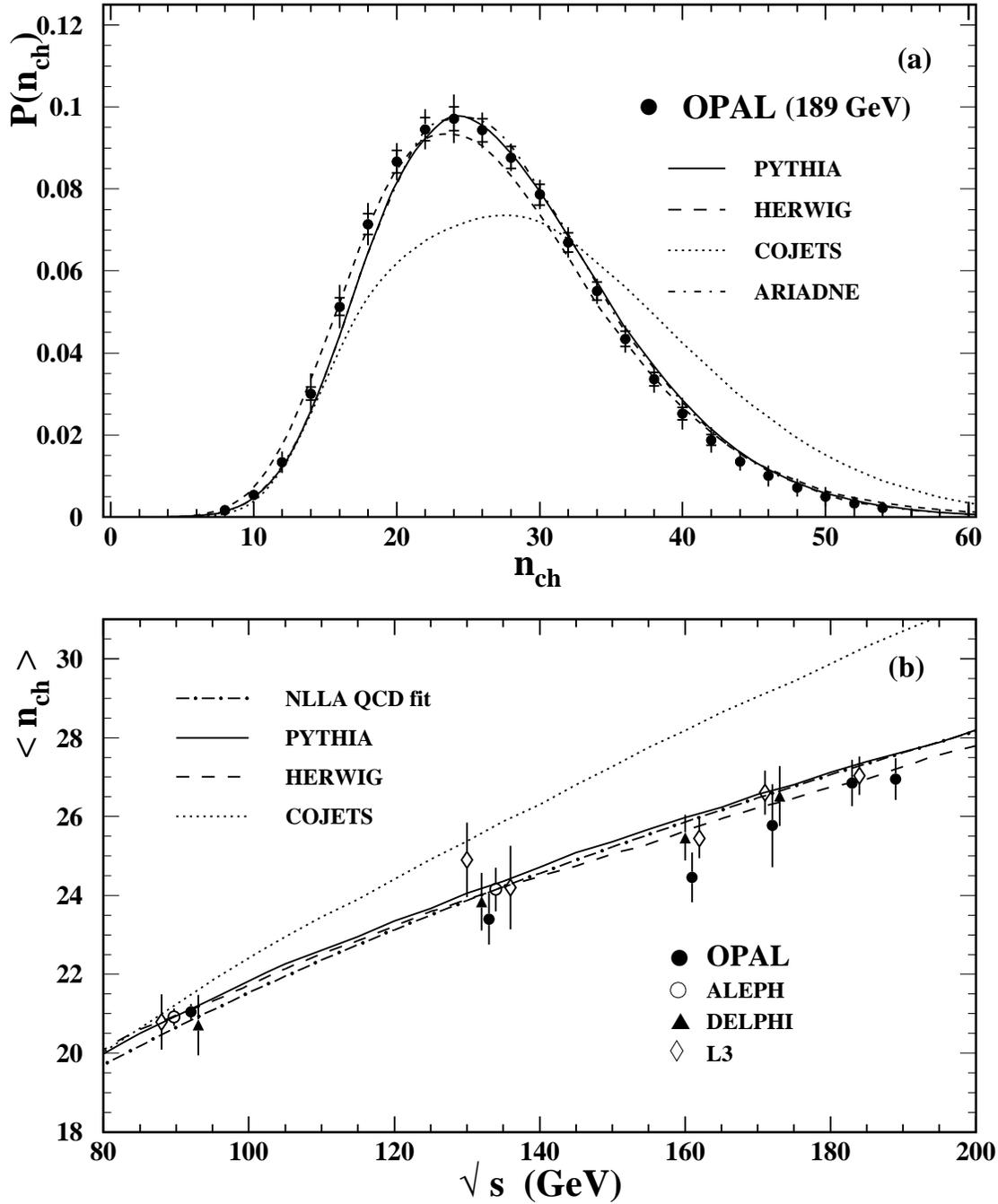}}
\caption[]
{ (a) Corrected distribution of the charged particle multiplicity
\nch.  (b) 
Mean charged multiplicity measurements by
LEP experiments over a range of \rs\ from 91.2 GeV to 189 GeV. 
The NLLA QCD prediction
for the evolution of charged particle
multiplicity with \rs\ 
uses all data points available from 12~GeV up to 161 GeV.
Also shown are the predictions  from PYTHIA, HERWIG,
 and COJETS. 
In (b) the curve for the ARIADNE prediction is
almost indistinguishable from the PYTHIA prediction and is omitted.
Note that the systematic uncertainties on the OPAL measurements at energies
$\rs > 91.2$ GeV are highly correlated.
}
\label{fig_nch}
\end{figure}

The charged particle multiplicity distribution, corrected for
experimental effects using an unfolding matrix, is tabulated in
Table~\ref{tab_ch}. The data are shown for our 189~GeV sample in
Figure~\ref{fig_nch}a.  The COJETS model, as indicated in the figure,
predicts too many high multiplicity events and clearly disagrees with
the data.  
The PYTHIA and ARIADNE models predict mutually indistinguishable distributions
which describe the data reasonably well. The HERWIG model gives a
somewhat better description for low and intermediate values.

We determine the mean values to be:
\begin{eqnarray} \nonumber 
  \mnch(172~\mathrm{GeV})  & = & 25.77 \pm0.58 \stat\pm0.88 \syst, \\ 
  \mnch(183~\mathrm{GeV})  & = & 26.85 \pm0.27 \stat\pm0.52 \syst, \\ \nonumber
  \mnch(189~\mathrm{GeV})  & = & 26.95 \pm0.16 \stat\pm0.51 \syst, \nonumber
\end{eqnarray}
The systematic errors are detailed in Table~9.  The
values of the dispersion $D$, the ratio $\langle
n_{\mathrm{ch}}\rangle/D$, the normalised second moment $C_{2}$ and
the second binomial moment \rtwo\ can be found in Table~\ref{tab_ch}.

As a consistency check, the mean charged multiplicity is also computed by 
integrating the corrected rapidity distribution, and the fragmentation function, as
determined in the next section.  Both the rapidity distribution and
the fragmentation function give results in good agreement with the one from
the direct measurement presented in this section. 

Figure~\ref{fig_nch}b shows our measurements of \mnch\ together with
results from OPAL at lower c.m.\ energies. The data are compared to
those of the other LEP experiments and to analytic QCD or Monte Carlo 
predictions.
The predicted value of
$\mnch(189~\mathrm{GeV})$ from PYTHIA
is  27.6. This value changes by up to 0.4 when the hadronization parameters 
$b$ and $\sigma_Q$ and the parton virtuality cutoff parameter $Q_0$ are varied by 
$\pm$~1 standard deviation about their tuned values~\cite{OPALPR141}. 
The values from ARIADNE and HERWIG are 27.5 and 27.2 respectively.
The measured value is therefore about
one standard deviations low compared to the PYTHIA and
ARIADNE models.  This trend is seen at all c.m.\ energies of
the LEP-2 data. However, this might be only a consequence of the
high correlation between the systematic uncertainties of the OPAL 
measurements. COJETS predicts the charged multiplicity to be
above 30, well above the measured value. The models have been 
tuned to agree with the OPAL data taken at the \znull\ peak with 
$\mnch(\mz)=21.05\pm0.01\stat\pm0.20\syst$~\cite{OPALPR133}.

The dispersion $D$ of the data,
$D(189~\mbox{GeV})=8.45\pm0.12\stat\pm0.34\syst$ is better described
by the PYTHIA and ARIADNE Monte Carlo samples, which predict 8.65 and 8.58
respectively. In contrast, the HERWIG model predicts $D(189~\mbox{GeV})=
9.25$, more than two standard deviations too large.

The dash-dotted curve in Figure~\ref{fig_nch}b shows the NLLA QCD 
prediction~\cite{webber84}
for the energy evolution of the charged particle multiplicity, with
parameters fitted to all available data points between 12 and
161~GeV~\cite{OPALPR158,nch_data,alephas133,l3as133,l3as161172,delphinch133,OPALPR133}.
The NLLA calculation predicts a mean charged multiplicity at $189$~GeV
of $27.6$.  Variations in the quark flavour composition at the
different c.m.\ energies do not significantly influence the results of
the fit.

\subsection{ Charged particle momentum spectra }
\label{sec_fragfun}
We measure the charged particle fragmentation function, $\sdscd\xp$,
and the \ksip\ distribution, $\sdscd\ksip$, where $\xp=2p/\rs$,
$\ksip=\ln(1/\xp)$ and $p$ is the measured track momentum, using the
same methods as in our previous publications~\cite{OPALPR158,OPALPR197}. We 
determine the rapidity distribution, $y= |\ln(\frac{E+p_\Vert}{E-p_\Vert})|$, where
 $p_\Vert$ is the momentum component
parallel to the thrust axis and $E$ is the energy of the particle.
We also study the distribution of the 3-momentum components parallel,
\ptin, and perpendicular, \ptout, to the event plane. This plane is
defined by the eigenvectors of the momentum tensor associated with the
two largest eigenvalues, as in reference~\cite{OPALPR158}.

\begin{figure}[!htb]
\resizebox{\textwidth}{!}
{\includegraphics{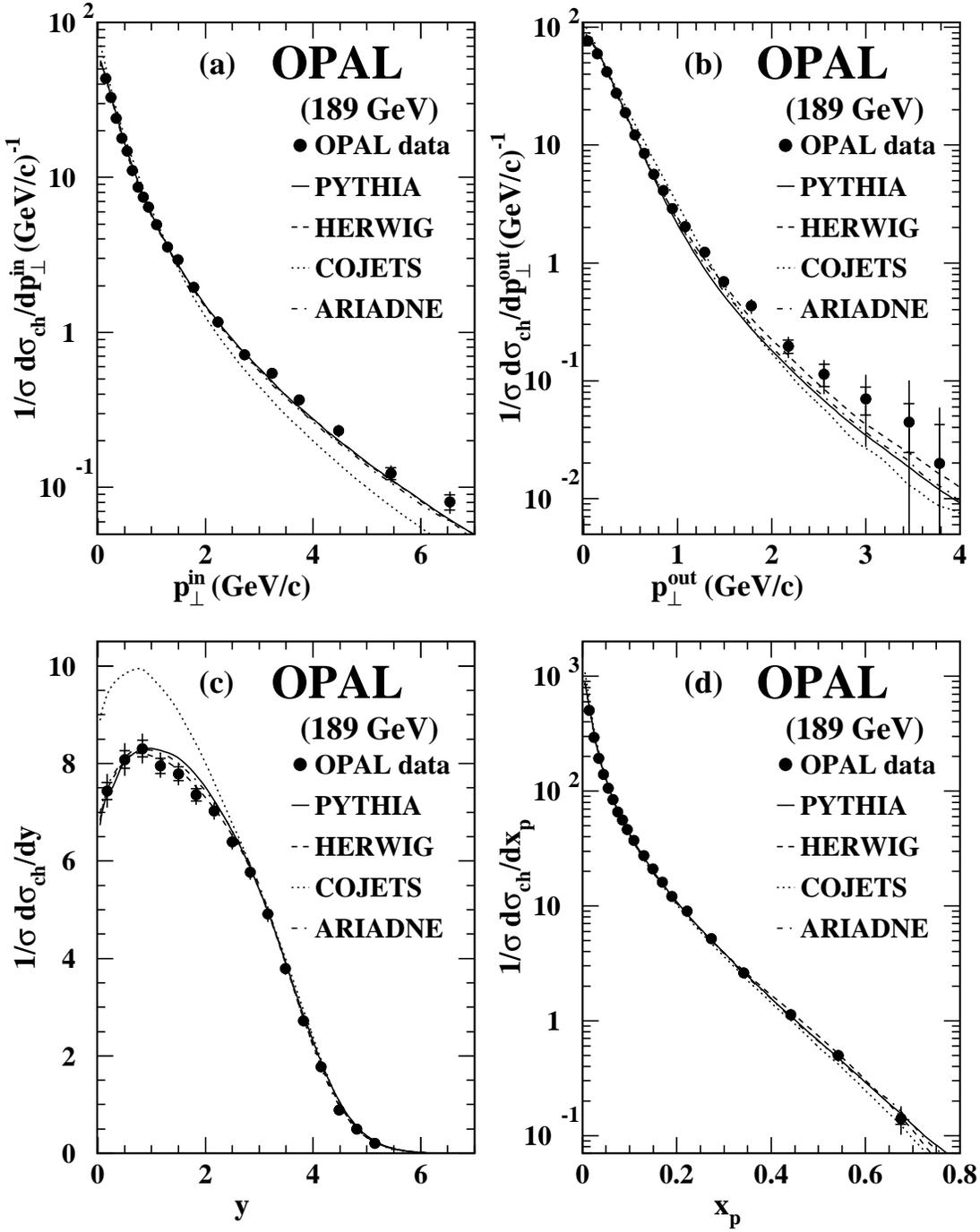}}
\caption[The \yp\ and \xp\ distributions]
{ (a,b,c) Distributions of the momenta \ptin, \ptout\ and of the rapidity, \yp;
(d) shows the fragmentation function $\sdscd\xp$ with
 $\xp=2p/\rs$; all compared to
 PYTHIA, HERWIG, COJETS and ARIADNE predictions. 
}
\label{fig_xpyp}
\end{figure}

The \ptin, \ptout\ and $y$ distributions for events with a c.m.\ energy of 
$\sqrt{s}=189$ GeV are shown in
Figure~\ref{fig_xpyp} and are tabulated in Tables~\ref{tab_ptin}, \ref{tab_ptout}
and \ref{tab_Y}. We observe good agreement
between PYTHIA, HERWIG and ARIADNE and the data for the \ptin\
distributions.  However, the slopes of the \ptin\ distributions
of the COJETS prediction are somewhat steeper  than the data. In case of the
\ptout\ distribution, not only COJETS but to lesser extent
also PYTHIA, HERWIG and ARIADNE predict a slightly softer \ptout\ 
spectrum than the spectrum observed in the data. 
In the \yp\ distribution, there is reasonable agreement of the
PYTHIA, HERWIG and ARIADNE Monte Carlo models with the data.
The COJETS Monte
Carlo model significantly overestimates the production of charged particles with low
values of \yp. 

\begin{figure}[!htb]
\resizebox{\textwidth}{!}
{\includegraphics{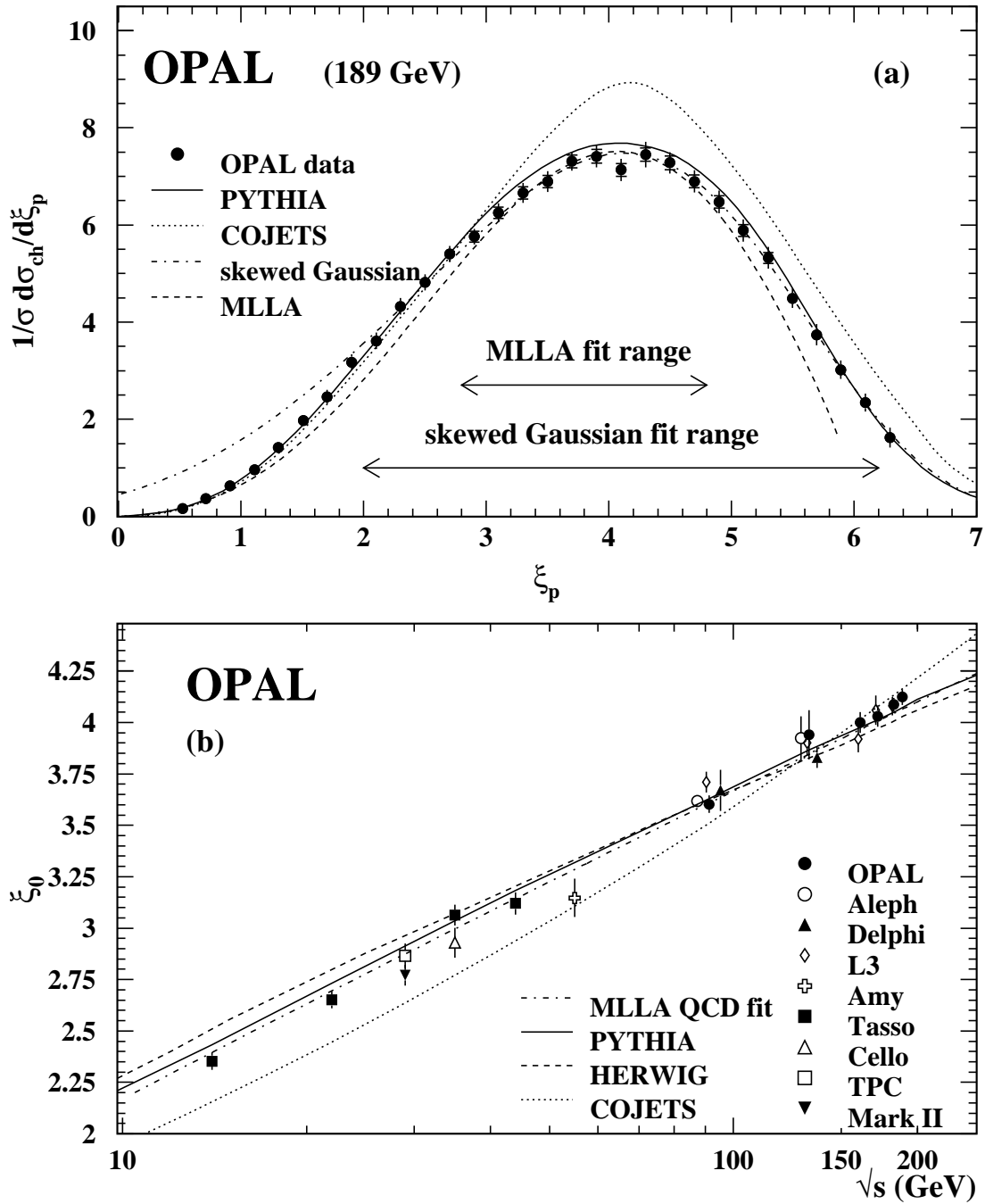}}
\caption[The \ksip\ distribution]
{ (a) Distribution of $\ksip=\ln(1/\xp)$ for charged particles at \rs = 189 GeV.
Also shown are a
fit of a skewed Gaussian and predictions by MLLA QCD, PYTHIA and COJETS. 
The curve for the ARIADNE prediction is
almost indistinguishable from the PYTHIA prediction and is omitted.
(b) Evolution of the position of the peak 
of the \ksip\ distribution, \ksinul, with c.m.\ energy \rs, compared with a fit
of a MLLA QCD prediction to the previous measurements and with predictions by PYTHIA, 
HERWIG and COJETS.}
\label{fig_ksip}
\end{figure}

The fragmentation function and the \ksip\ distribution are shown in
Figures~\ref{fig_xpyp}~(d) and~\ref{fig_ksip}~(a) together with the
Monte Carlo predictions.  Numerical values of these data are given in
Tables~\ref{tab_xp} and \ref{ksip_tab}.  The spectrum of charged
particles with large momentum fraction \xp\ is well described by all
Monte Carlo models. The shape of the \ksip\ distribution is well
described by PYTHIA, although their normalisation
is somewhat higher, reflecting the difference between the predicted and the
observed charged particle multiplicity (see Section~\ref{sec_mult}).  
The COJETS Monte Carlo predicts 
too many particles in the region of the peak and at large values of \ksip,
where low momentum particles contribute.  In the LLA approach, this is
the region where soft gluon production is reduced as a consequence of
destructive interference.

Based on the concept of local parton hadron duality (LPHD)\cite{lphd} within the
leading-log approximation of QCD calculations (LLA)\cite{dfk} one expects a
Gaussian shape of the \ksip\ distribution.  The peak of the
distribution is predicted to have an almost logarithmic variation with
energy.  The next to leading-log approximation of  QCD 
calculations\cite{lphd,fong/webber91} 
predicts the shape to be a Gaussian with higher moments (skewed Gaussian), and the
same energy dependence of the peak as the LLA.  We fit a skewed
Gaussian of the form suggested in \cite{fong/webber91}
to the region $2.0 < \ksip < 6.2$ of the three \ksip\ distributions
and determine the positions of the peaks, \ksinul, to be
\begin{eqnarray}
  \ksinul(172~\mathrm{GeV}) & = & 4.031\pm0.033\stat\pm0.041\syst,\nonumber \\
  \ksinul(183~\mathrm{GeV}) & = & 4.087\pm0.014\stat\pm0.030\syst, \\
  \ksinul(189~\mathrm{GeV}) & = & 4.124\pm0.010\stat\pm0.037\syst.\nonumber
\end{eqnarray}
The systematic uncertainty takes into account experimental effects as described in
Section~\ref{sec_syserr} and the uncertainty due to the choice of the fit range. The fit range was
reduced to $3.2 < \ksip < 4.8$ and increased to $1.6 < \ksip < 6.6$. 
The larger deviation from the result with the standard fit range was taken as systematic uncertainty
and added in quadrature to the other effects. All systematic effects are summarised in 
Table 15.
  
In Figure~\ref{fig_ksip}~(b) we show our measurements of \ksinul\ together
with measurements taken at various c.m.\
energies~\cite{OPALPR197,OPALPR158,alephas133,delphias133,l3as183,OPALPR017,peak_data},
and the PYTHIA, HERWIG and COJETS predictions. The dash-dotted curve shows the
prediction for the energy evolution by the modified leading-log approximation (MLLA) 
\cite{modmlla2} with parameters fitted to the different energies, 
excluding the results of this paper. The values of the MLLA prediction extrapolated to
$\rs=172$, 183 and 189~GeV of $4.01\pm0.01$, $4.05\pm0.01$ and $4.07\pm0.01$ 
are lower than our measurements.  As noted in our previous
publications~\cite{OPALPR158,OPALPR197}, the fit is dominated by contributions
from the data points with small errors at 29 and 35~GeV. The PYTHIA and HERWIG
predictions are very similar to the MLLA fit, while the COJETS prediction 
does not agree with the data.

\begin{figure}[!htb]
\begin{picture}(150,103)
\resizebox{\textwidth}{!}
{\includegraphics{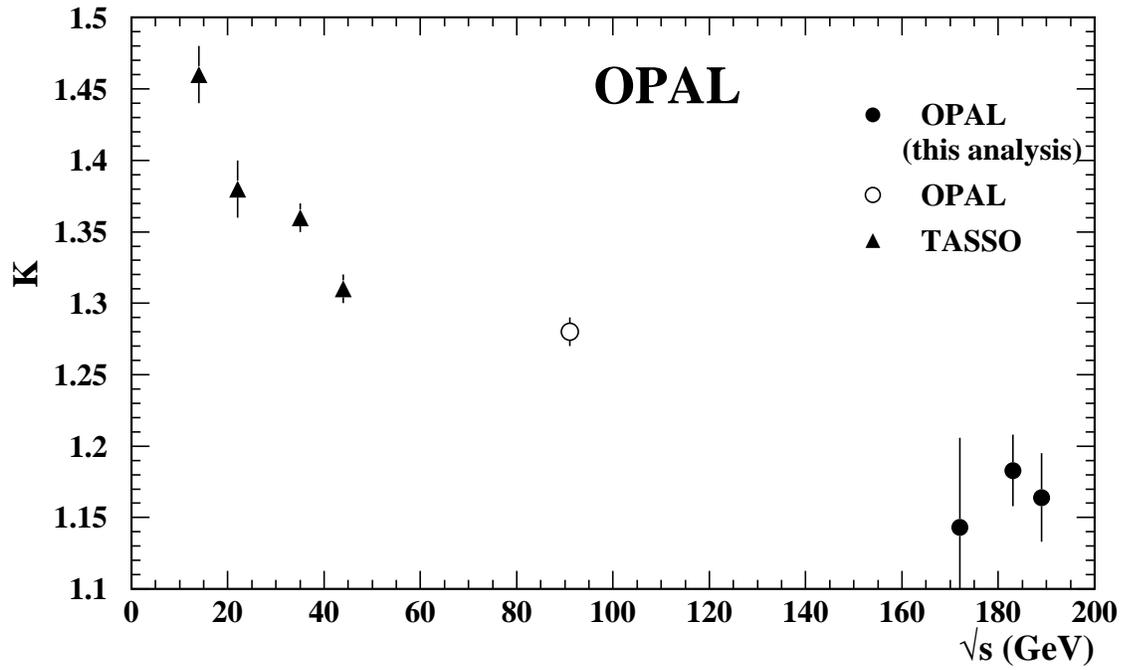}}
\end{picture}
\caption[ The MLLA normalisation factor $K$]
{Results for the normalisation factor $K$ in the MLLA description of
     the \ksip\ distribution, compared with results from fits to the \ksip\
     distribution at lower energies ~\cite{OPALPR017}.}
\label{fig_kmlla}
\end{figure}

In Figure ~\ref{fig_ksip}~(a) also the shape predicted by MLLA~\cite{mlla} is shown. 
The MLLA prediction is determined by two parameters, an effective QCD scale 
$\Lambda_{eff}$ and a normalisation factor $K(\rs)$ 
which within the framework of LPHD is expected to have a weak c.m. energy dependence.  
We fix $ \Lambda  _{eff} $ to the value determined 
in ~\cite{OPALPR017}, $ \Lambda  _{eff} =0.253 $, and fit the formula to the measured 
charged particle momentum spectra in the region $2.8 < \ksip < 4.8$. The result of this
fit to the $\rs=$189~GeV data is shown in Figure ~\ref{fig_ksip}~(a).
The MLLA description and data show good agreement in the peak region but
disagree at large
\ksip\ where kinematic effects become important and
the perturbative QCD calculations are not valid \cite{pertQCDnotvalid}. 
The values obtained for the normalisation factor $K$ are:
\begin{eqnarray}
  K(\rs=172~\mathrm{GeV}) & = & 1.143\pm0.030\stat\pm0.056\syst,\nonumber \\
  K(\rs=183~\mathrm{GeV}) & = & 1.183\pm0.010\stat\pm0.023\syst, \\
  K(\rs=189~\mathrm{GeV}) & = & 1.164\pm0.008\stat\pm0.030\syst.\nonumber
\end{eqnarray}
To estimate the uncertainty due to the choice of the fit range, the fit range was
reduced to $3.6 < \ksip < 4.2$ and increased to $2.4 < \ksip < 5.2$. 
the systematic uncertainty is shown in Table~16. In Figure~\ref{fig_kmlla}, these
numbers are compared with results for the normalisation factors obtained in~\cite{OPALPR017}
 for $\ksip$ distributions at lower c.m.~energies. 
The values from the new measurements are significantly below the value at 
 $\sqrt{s}= \mz$, $K(\mz) = 1.28 \pm 0.01$, thus following the observed 
trend of decreasing values for $K$ with increasing c.m.~energy. A decrease of $K$ values might be 
caused by higher order effects as discussed in \cite{kdecrease}.

MLLA makes several predictions for relations between the different quantities that
can be used to describe the centre of the \ksip\ distribution, i.e.~the position of 
the maximum  \ksinul, the mean value \ksimean\ and the median
value \ksimed\ which is defined as the value where 
$\int_0^{\ksimed}(\mathrm{d}\sigma_{\mathrm{ch}}/\mathrm{d}\ksip) {\rm d}\ksip$
    is equal to 
$\int_{\ksimed}^{\infty}(\mathrm{d}\sigma_{\mathrm{ch}}/\mathrm{d}\ksip) 
{\rm d}\ksip$. The difference $\ksinul-\ksimean$ is 
predicted to be approximately 0.351 (0.355) for 3(5) active flavours,
independent from the c.m.~energy\cite{mlla}. 
The first non-leading corrections to this value cannot be predicted at the
moment. However, this correction has been taken into account for the calculation of 
the ratio of differences, $(\ksinul-\ksimean)/(\ksimed-\ksimean)$ which is 
predicted to be 3.5 (3.4) for 3(5) active flavours for the c.m~energies considered
in this paper and assuming $ \Lambda  _{eff} =0.253 $.
In our analysis,
 we measure
\begin{eqnarray}
  \ksinul-\ksimean(172~\mathrm{GeV}) & = &\mbox{\hspace{0.2cm} } 0.054\pm0.023\stat\pm0.063\syst,\nonumber \\
  \ksinul-\ksimean(183~\mathrm{GeV}) & = &\mbox{\hspace{0.2cm} } 0.012\pm0.010\stat\pm0.064\syst, \\
  \ksinul-\ksimean(189~\mathrm{GeV}) & = &                      -0.035\pm0.007\stat\pm0.075\syst\nonumber
\end{eqnarray}
and
\begin{eqnarray}
  (\ksinul-\ksimean)/(\ksimed-\ksimean)(172~\mathrm{GeV}) & =
  &-0.38\pm0.56\stat\pm0.42\syst,\nonumber \\
  (\ksinul-\ksimean)/(\ksimed-\ksimean)(183~\mathrm{GeV}) & = 
  &-0.06\pm0.44\stat\pm0.27\syst, \\
  (\ksinul-\ksimean)/(\ksimed-\ksimean)(189~\mathrm{GeV}) & = 
  &\mbox{\hspace{0.2cm} }0.16\pm0.33\stat\pm0.22\syst.\nonumber
\end{eqnarray}
The systematic errors are detailed in Tables~17 and 18. The results strongly depend
on the high \ksip\ region of the \ksip\ spectrum, i.e.~on the low momentum region
where the selection efficiency for tracks is low and the corrections large.
In MLLA mass effects are neglected and
the \ksip\ distribution vanishes for values above $\ksip = \log(E/\Lambda  _{eff})$. 
In \cite{modmlla1,modmlla2}, a simple procedure was suggested on how to take these
effects into account in the MLLA predictions.  Indeed, when we
fitted these `modified' MLLA predictions to the \ksip\ distribution
at $\rs = 189$ GeV, it turned out that the fitted values for \ksinul,
\ksimed\ and \ksimean\ became numerically very close to each
other such that the value for the difference $\ksinul-\ksimean$
almost vanished, consistent with our measurement.
This also indicates that the value and even the sign for the ratio
$(\ksinul-\ksimean)/(\ksimed-\ksimean)$ is largely undetermined, and
depends strongly on the assumptions made in the correction procedure
to the MLLA distribution. A comparison with the measured values is
therefore not conclusive.

\section{ Summary and conclusions }
\label{sec_conc}
In this paper we have presented measurements of the properties of
hadronic events produced at LEP at centre-of-mass energies between 172
and 189~GeV. The 189~GeV data-set recorded by
the OPAL detector constitutes approximately 70\perc\ of the total luminosity
of the LEP-2 programme available by the end of 1998, and provides the most precise
results at energies above the \zzero\ resonance.

We have determined the corrected distributions for event shape
observables, for the charged particle multiplicity and for charged
particle momentum spectra at
centre-of-mass energies of $\rs=$172, 183 and 189~GeV.  The predictions
of the PYTHIA, HERWIG and ARIADNE Monte Carlo models are found to be
in general agreement with the measured distributions.
While the COJETS Monte Carlo model describes most event shape distributions
well, it overestimates the number of soft charged particles produced as
observed in the rapidity and $\ksip=\ln(1/\xp)$ distributions.

From a fit of \oaa+NLLA QCD predictions to five event shapes
 and the differential two jet rate, defined
using the Durham jet finder, we have determined the strong coupling
parameter $\as(187~\mathrm{GeV})=0.106\pm0.001\stat\pm
0.004\syst$ at a luminosity weighted centre-of-mass energy of 186.7 GeV.
When this is evolved to the \znull\ peak it is in excellent agreement with 
our previous measurement at $\rs=\mz$. 

The mean charged particle multiplicity has been determined to be
$\mnch(189~\mathrm{GeV})=26.95\pm0.16\stat\pm0.51\syst$, 
about one standard deviations below the NLLA QCD predictions and
the PYTHIA and ARIADNE  
Monte Carlo predictions.

The \ksip\ distribution is not in perfect agreement with 
QCD  MLLA calculations. The position of the peak was
measured to be
$\ksinul(189~\mathrm{GeV})=4.124\pm0.010\stat\pm0.037\syst$. This is
approximately one and a half standard deviations larger than the expectation for
the energy evolution given by a QCD MLLA extrapolation from 
the results of lower energy experiments.

Our studies show that most of the features of hadronic events
produced in \epem\ collisions at energies
above the \znull\ mass are well described by QCD in the form
of analytic or Monte Carlo predictions, or both. 
\par
\section*{Acknowledgements}
\par
We particularly wish to thank the SL Division for the efficient operation
of the LEP accelerator at all energies
 and for their continuing close cooperation with
our experimental group.  We thank our colleagues from CEA, DAPNIA/SPP,
CE-Saclay for their efforts over the years on the time-of-flight and trigger
systems which we continue to use.  In addition to the support staff at our own
institutions we are pleased to acknowledge the  \\
Department of Energy, USA, \\
National Science Foundation, USA, \\
Particle Physics and Astronomy Research Council, UK, \\
Natural Sciences and Engineering Research Council, Canada, \\
Israel Science Foundation, administered by the Israel
Academy of Science and Humanities, \\
Minerva Gesellschaft, \\
Benoziyo Center for High Energy Physics,\\
Japanese Ministry of Education, Science and Culture (the
Monbusho) and a grant under the Monbusho International
Science Research Program,\\
Japanese Society for the Promotion of Science (JSPS),\\
German Israeli Bi-national Science Foundation (GIF), \\
Bundesministerium f\"ur Bildung, Wissenschaft,
Forschung und Technologie, Germany, \\
National Research Council of Canada, \\
Research Corporation, USA,\\
Hungarian Foundation for Scientific Research, OTKA T-029328, 
T023793 and OTKA F-023259.\\

%


\begin{table}[!htb]
  \begin{center}
    \begin{tabular}{|lr|ccc|} \hline 
                    &                  & 172 GeV             & 183 GeV         & 189 GeV      \\ \hline\hline
      $\langle$c.m.\ energy$\rangle$   & (GeV)              & 172.1               & 182.7           & 188.7        \\ 
      Luminosity    & (pb$^{-1}$)      & 10.4                &  57.2           & 186.3        \\ \hline
      Preselection  &                  &  1318               & 6142            &  18954       \\ \hline
{\bf `ISR-fit' Selection}  &                  & {\bf 273}           & {\bf 1447}      &{\bf  4457 }  \\
      4-fermion background      &\perc & 20.3                & 26.6            & 29.4         \\
      Efficiency non-rad events &\perc & 82.6                & 83.3            & 82.9         \\ \hline
{\bf Final selection}           &      & {\bf 228}           & {\bf 1098}      &{\bf 3277  }  \\ 
      4-fermion background      &\perc & 7.3                 &  9.0            & 10.1         \\
      Efficiency non-rad events &\perc & 79.4                &  79.4           & 78.5         \\ \hline 
    \end{tabular}
  \end{center}
  \caption[]{
    Data samples at 172, 183 and 189~GeV. The luminosity and
luminosity weighted mean c.m.\ energies are given in the first two
rows. The rows labelled as `ISR-fit selection' and `Final
selection' correspond to the number of events passing these criteria.
}
  \label{t:select}
\end{table}

\begin{table}[!ht]
  \begin{center}
    {\small
      \begin{tabular}{|c|r@{ $\pm$ }l@{ $\pm$ }l|
          r@{ $\pm$ }l@{ $\pm$ }l|
          r@{ $\pm$ }l@{ $\pm$ }l|} 
        \hline
        $T$  & \multicolumn{3}{c||}{ $R(T)$  (172 GeV) }
        &      \multicolumn{3}{c||}{ $R(T)$  (183 GeV) }
        &      \multicolumn{3}{c||}{ $R(T)$  (189 GeV) } \\  
        \hline\hline
0.700-0.780
     &   0.62 &   0.27 &   0.31
     &   0.29 &   0.11 &   0.17
     &   0.36 &   0.09 &   0.14\\
0.780-0.850
     &   0.60 &   0.28 &   0.22
     &   0.75 &   0.15 &   0.16
     &   0.78 &   0.09 &   0.14\\
0.850-0.880
     &   0.73 &   0.45 &   0.54
     &   1.45 &   0.28 &   0.44
     &   1.53 &   0.17 &   0.22\\
0.880-0.910
     &   1.35 &   0.52 &   1.77
     &   1.62 &   0.27 &   0.68
     &   2.22 &   0.18 &   0.16\\
0.910-0.930
     &   3.01 &   0.91 &   2.91
     &   3.30 &   0.44 &   0.60
     &   3.09 &   0.25 &   0.45\\
0.930-0.950
     &   6.60 &   1.25 &   1.77
     &   5.27 &   0.54 &   0.64
     &   5.14 &   0.31 &   0.46\\
0.950-0.960
     &   4.47 &   1.43 &   3.20
     &   8.20 &   0.95 &   0.95
     &   7.02 &   0.50 &   0.88\\
0.960-0.970
     &   7.85 &   1.90 &   3.19
     &   10.0 &    1.0 &    1.4
     &   9.94 &   0.59 &   0.56\\
0.970-0.980
     &   18.5 &    2.9 &    6.8
     &   15.1 &    1.3 &    0.8
     &   15.6 &    0.7 &    1.2\\
0.980-0.990
     &   22.3 &    3.4 &    3.7
     &   22.9 &    1.7 &    1.6
     &   22.0 &    0.9 &    1.6\\
0.990-1.000
     &   11.6 &    2.6 &    3.8
     &   9.08 &   1.07 &   1.46
     &   8.73 &   0.60 &   0.63\\
        \hline
        \tma  & \multicolumn{3}{c||}{ $R(\tma)$  (172 GeV) }
        & \multicolumn{3}{c||}{ $R(\tma)$  (183 GeV) }
        & \multicolumn{3}{c||}{ $R(\tma)$  (189 GeV) } \\  
        \hline\hline
0.000-0.040
     &   1.68 &   0.57 &   0.62
     &   0.92 &   0.19 &   0.16
     &   0.88 &   0.11 &   0.14\\
0.040-0.080
     &   7.33 &   0.99 &   0.68
     &   7.17 &   0.47 &   0.34
     &   6.66 &   0.26 &   0.35\\
0.080-0.120
     &   4.87 &   0.72 &   0.18
     &   4.96 &   0.36 &   0.27
     &   5.00 &   0.21 &   0.22\\
0.120-0.160
     &   3.23 &   0.60 &   1.09
     &   3.22 &   0.29 &   0.60
     &   3.19 &   0.17 &   0.33\\
0.160-0.220
     &   2.03 &   0.40 &   0.28
     &   2.32 &   0.20 &   0.19
     &   2.36 &   0.12 &   0.20\\
0.220-0.300
     &   1.07 &   0.27 &   0.38
     &   1.18 &   0.14 &   0.24
     &   1.26 &   0.08 &   0.14\\
0.300-0.400
     &   0.49 &   0.21 &   0.21
     &   0.84 &   0.12 &   0.21
     &   0.73 &   0.07 &   0.05\\
0.400-0.500
     &   0.29 &   0.18 &   0.13
     &   0.23 &   0.08 &   0.12
     &   0.35 &   0.06 &   0.12\\
0.500-0.600
     &   0.33 &   0.17 &   0.24
     &  0.033 &  0.075 &  0.124
     &   0.18 &   0.06 &   0.11\\
        \hline
        \tmi & \multicolumn{3}{c||}{ $R(\tmi)$  (172 GeV) }
        & \multicolumn{3}{c||}{ $R(\tmi)$  (183 GeV) }
        & \multicolumn{3}{c||}{ $R(\tmi)$  (189 GeV) } \\  
        \hline\hline
0.000-0.020
     &   0.59 &   0.43 &   0.54
     &  0.087 &  0.067 &  0.129
     &   0.22 &   0.06 &   0.12\\
0.020-0.040
     &   10.6 &    1.8 &    1.5
     &   9.49 &   0.80 &   0.82
     &   8.03 &   0.41 &   1.38\\
0.040-0.060
     &   16.8 &    2.1 &    5.6
     &   15.7 &    1.0 &    1.5
     &   14.7 &    0.5 &    1.7\\
0.060-0.080
     &   7.73 &   1.31 &   2.77
     &   10.6 &    0.8 &    1.8
     &   10.1 &    0.4 &    1.3\\
0.080-0.100
     &   6.24 &   1.16 &   2.16
     &   5.86 &   0.55 &   1.26
     &   5.54 &   0.31 &   0.50\\
0.100-0.120
     &   4.07 &   0.99 &   2.07
     &   3.46 &   0.44 &   0.72
     &   2.95 &   0.24 &   0.24\\
0.120-0.140
     &   1.52 &   0.66 &   1.45
     &   2.69 &   0.44 &   0.86
     &   2.06 &   0.22 &   0.32\\
0.140-0.160
     &   1.16 &   0.67 &   0.97
     &   1.24 &   0.33 &   0.26
     &   1.55 &   0.21 &   0.36\\
0.160-0.200
     &   0.48 &   0.39 &   0.77
     &   0.69 &   0.23 &   0.66
     &   0.69 &   0.14 &   0.44\\
0.200-0.240
     &   0.16 &   0.32 &   0.76
     &   0.20 &   0.22 &   0.55
     &   0.12 &   0.15 &   0.75\\
0.240-0.300
     &   0.30 &   0.48 &   0.86
     &  -.475 &  0.617 &  1.540
     &   0.15 &   0.18 &   0.46\\
        \hline
        $A$ & \multicolumn{3}{c||}{ $R(A)$  (172 GeV) }
        & \multicolumn{3}{c||}{ $R(A)$  (183 GeV) }
        & \multicolumn{3}{c||}{ $R(A)$  (189 GeV) } \\  
        \hline\hline
0.000-0.005
     &   119. &    10. &    34.
     &   131. &     5. &     5.
     &   129. &     3. &     9.\\
0.005-0.010
     &   33.2 &    5.9 &    9.8
     &   34.2 &    3.1 &    3.3
     &   30.9 &    1.7 &    1.8\\
0.010-0.015
     &   14.2 &    4.0 &    9.4
     &   12.6 &    2.0 &    4.4
     &   12.7 &    1.2 &    0.8\\
0.015-0.025
     &   5.88 &   2.15 &   2.02
     &   5.56 &   1.07 &   2.28
     &   4.87 &   0.62 &   1.40\\
0.025-0.040
     &   0.80 &   0.91 &   1.42
     &   2.55 &   0.72 &   1.26
     &   2.17 &   0.44 &   0.58\\
0.040-0.070
     &  -.442 &  0.336 &  0.400
     &   0.58 &   0.44 &   1.42
     &   0.59 &   0.25 &   0.43\\
0.070-0.100
     &   0.14 &   0.44 &   7.30
     &  -.419 &  0.620 &  1.281
     &   0.14 &   0.32 &   0.66\\
        \hline  
      \end{tabular}
      }
  \end{center}
  \caption[]
  { Results $R(y)=\sdsd y$ at $\rs=172$, 183 and 189~GeV for the event
    shape observables $y$: thrust $T$, thrust major \tma, thrust minor \tmi\
    and aplanarity $A$. The first error is statistical, the second systematic.
    }
  \label{t:evsset1}
\end{table}

\begin{table}[!ht]
  \begin{center}
    {\small 
      \begin{tabular}{|c|r@{ $\pm$ }l@{ $\pm$ }l|
          r@{ $\pm$ }l@{ $\pm$ }l|
          r@{ $\pm$ }l@{ $\pm$ }l|} 
        \hline
        $O$  & \multicolumn{3}{c||}{ $R(O)$  (172 GeV) }
        & \multicolumn{3}{c||}{ $R(O)$  (183 GeV) }
        & \multicolumn{3}{c||}{ $R(O)$  (189 GeV) } \\  
        \hline\hline
0.000-0.050
     &   10.6 &    1.0 &    0.5
     &   10.2 &    0.5 &    0.4
     &   9.97 &   0.27 &   0.44\\
0.050-0.100
     &   4.63 &   0.68 &   0.71
     &   4.30 &   0.31 &   0.51
     &   4.15 &   0.18 &   0.38\\
0.100-0.150
     &   1.93 &   0.47 &   0.23
     &   2.18 &   0.23 &   0.25
     &   2.33 &   0.14 &   0.11\\
0.150-0.200
     &   0.91 &   0.34 &   0.23
     &   1.32 &   0.20 &   0.57
     &   1.34 &   0.12 &   0.23\\
0.200-0.250
     &   0.46 &   0.30 &   0.37
     &   0.78 &   0.16 &   0.47
     &   0.82 &   0.10 &   0.21\\
0.250-0.300
     &   0.51 &   0.29 &   0.19
     &   0.72 &   0.16 &   0.19
     &   0.60 &   0.09 &   0.18\\
0.300-0.400
     &   0.28 &   0.15 &   0.21
     &   0.21 &   0.07 &   0.09
     &   0.29 &   0.05 &   0.10\\
0.400-0.500
     &   0.21 &   0.11 &   0.15
     &  0.031 &  0.034 &  0.062
     &  0.087 &  0.028 &  0.057\\
        \hline
        \cp  & \multicolumn{3}{c||}{ $R(\cp)$  (172 GeV) }
        & \multicolumn{3}{c||}{ $R(\cp)$  (183 GeV) }
        & \multicolumn{3}{c||}{ $R(\cp)$  (189 GeV) } \\  
        \hline\hline
0.000-0.050
     &   2.76 &   0.59 &   1.12
     &   1.78 &   0.22 &   0.33
     &   1.82 &   0.13 &   0.17\\
0.050-0.080
     &   4.65 &   0.92 &   2.48
     &   5.70 &   0.49 &   0.38
     &   4.97 &   0.26 &   0.19\\
0.080-0.110
     &   4.68 &   0.86 &   2.48
     &   3.85 &   0.37 &   0.60
     &   4.21 &   0.22 &   0.67\\
0.110-0.140
     &   3.66 &   0.73 &   0.86
     &   3.31 &   0.34 &   0.43
     &   3.14 &   0.19 &   0.47\\
0.140-0.180
     &   2.12 &   0.48 &   0.86
     &   2.28 &   0.24 &   0.33
     &   2.45 &   0.15 &   0.24\\
0.180-0.220
     &   1.31 &   0.39 &   0.48
     &   1.77 &   0.21 &   0.25
     &   1.80 &   0.13 &   0.28\\
0.220-0.300
     &   1.62 &   0.31 &   0.35
     &   1.46 &   0.14 &   0.13
     &   1.27 &   0.08 &   0.13\\
0.300-0.400
     &   0.82 &   0.21 &   0.21
     &   0.80 &   0.09 &   0.10
     &   0.86 &   0.06 &   0.08\\
0.400-0.500
     &   0.28 &   0.14 &   0.38
     &   0.60 &   0.09 &   0.20
     &   0.66 &   0.05 &   0.08\\
0.500-0.600
     &   0.28 &   0.15 &   0.26
     &   0.44 &   0.08 &   0.19
     &   0.42 &   0.05 &   0.14\\
0.600-0.750
     &   0.45 &   0.16 &   0.24
     &   0.26 &   0.07 &   0.09
     &   0.35 &   0.05 &   0.07\\
0.750-1.000
     &  -.002 &  0.057 &  0.153
     &  0.093 &  0.058 &  0.102
     &  0.082 &  0.035 &  0.060\\
        \hline
        \mh & \multicolumn{3}{c||}{ $R(\mh)$  (172 GeV) }
        & \multicolumn{3}{c||}{ $R(\mh)$  (183 GeV) }
        & \multicolumn{3}{c||}{ $R(\mh)$  (189 GeV) } \\  
        \hline\hline
0.060-0.075
     &   1.57 &   0.53 &   1.32
     &   1.20 &   0.23 &   0.47
     &   1.67 &   0.17 &   0.23\\
0.075-0.090
     &   5.52 &   1.26 &   1.42
     &   4.95 &   0.57 &   0.98
     &   4.35 &   0.31 &   0.48\\
0.090-0.110
     &   7.15 &   1.37 &   2.60
     &   6.31 &   0.61 &   0.88
     &   6.07 &   0.34 &   0.56\\
0.110-0.140
     &   5.84 &   0.99 &   1.69
     &   5.62 &   0.48 &   0.63
     &   6.42 &   0.29 &   0.39\\
0.140-0.170
     &   4.27 &   0.84 &   1.44
     &   5.06 &   0.44 &   0.80
     &   4.77 &   0.25 &   0.35\\
0.170-0.200
     &   4.03 &   0.84 &   1.18
     &   3.88 &   0.39 &   0.66
     &   3.22 &   0.20 &   0.32\\
0.200-0.250
     &   2.70 &   0.54 &   0.50
     &   2.71 &   0.25 &   0.42
     &   2.45 &   0.14 &   0.15\\
0.250-0.300
     &   0.88 &   0.36 &   0.64
     &   1.56 &   0.21 &   0.14
     &   1.66 &   0.13 &   0.11\\
0.300-0.350
     &   0.72 &   0.34 &   0.21
     &   1.08 &   0.19 &   0.32
     &   1.37 &   0.13 &   0.19\\
0.350-0.450
     &   0.56 &   0.22 &   0.09
     &   0.57 &   0.11 &   0.04
     &   0.57 &   0.07 &   0.14\\
0.450-0.600
     &   0.34 &   0.13 &   0.19
     &   0.11 &   0.05 &   0.07
     &   0.14 &   0.03 &   0.10\\
        \hline  
        \hline
        $S$ & \multicolumn{3}{c||}{ $R(S)$  (172 GeV) }
        & \multicolumn{3}{c||}{ $R(S)$  (183 GeV) }
        & \multicolumn{3}{c||}{ $R(S)$  (189 GeV) } \\  
        \hline\hline
0.000-0.020
     &   26.7 &    2.5 &    2.0
     &   24.7 &    1.1 &    2.0
     &   25.0 &    0.7 &    0.8\\
0.020-0.040
     &   8.19 &   1.43 &   1.60
     &   7.70 &   0.65 &   1.16
     &   7.46 &   0.38 &   0.69\\
0.040-0.060
     &   3.63 &   0.95 &   0.60
     &   4.18 &   0.51 &   0.71
     &   4.15 &   0.29 &   0.49\\
0.060-0.120
     &   1.43 &   0.39 &   0.26
     &   2.04 &   0.21 &   0.16
     &   1.90 &   0.12 &   0.12\\
0.120-0.200
     &   0.60 &   0.25 &   0.41
     &   0.65 &   0.12 &   0.17
     &   0.79 &   0.08 &   0.18\\
0.200-0.300
     &   0.24 &   0.19 &   0.22
     &   0.44 &   0.11 &   0.22
     &   0.37 &   0.07 &   0.04\\
0.300-0.500
     &   0.17 &   0.11 &   0.16
     &  0.093 &  0.060 &  0.093
     &   0.22 &   0.05 &   0.07\\
0.500-0.700
     &   0.19 &   0.12 &   0.25
     &   0.16 &   0.08 &   0.13
     &  0.055 &  0.030 &  0.046\\
        \hline  
      \end{tabular}
      }
  \end{center}
  \caption[] 
  { Results $R(y)=\sdsd y$ at $\rs=172$, 183 and 189~GeV for the event
    shape observables $y$: oblateness $O$, C-parameter \cp, heavy jet mass
    \mh\ and sphericity $S$. The first error is statistical, the second systematic.
    }
  \label{t:evsset2}
\end{table}

\begin{table}[!ht]
  \begin{center}
    {\small 
      \begin{tabular}{|c|r@{ $\pm$ }l@{ $\pm$ }l|
          r@{ $\pm$ }l@{ $\pm$ }l|
          r@{ $\pm$ }l@{ $\pm$ }l|} 
        \hline
        \bt  & \multicolumn{3}{c||}{ $R(\bt)$  (172 GeV) }
        & \multicolumn{3}{c||}{ $R(\bt)$  (183 GeV) }
        & \multicolumn{3}{c||}{ $R(\bt)$  (189 GeV) } \\  
        \hline\hline
0.000-0.030
     &   3.96 &   0.98 &   1.68
     &   1.40 &   0.26 &   0.60
     &   1.89 &   0.17 &   0.36\\
0.030-0.040
     &   11.3 &    2.6 &    2.6
     &   13.4 &    1.3 &    2.0
     &   9.84 &   0.64 &   0.56\\
0.040-0.050
     &   8.75 &   2.09 &   7.73
     &   12.9 &    1.3 &    2.7
     &   11.9 &    0.7 &    1.7\\
0.050-0.060
     &   11.7 &    2.3 &    9.7
     &   8.57 &   0.91 &   1.54
     &   9.59 &   0.57 &   0.46\\
0.060-0.075
     &   8.82 &   1.59 &   2.61
     &   8.69 &   0.78 &   0.96
     &   8.00 &   0.42 &   0.53\\
0.075-0.090
     &   4.46 &   1.14 &   3.65
     &   5.20 &   0.58 &   0.99
     &   6.33 &   0.38 &   0.55\\
0.090-0.110
     &   5.53 &   1.12 &   1.06
     &   5.67 &   0.56 &   0.91
     &   4.37 &   0.27 &   0.61\\
0.110-0.130
     &   3.19 &   0.85 &   1.47
     &   3.11 &   0.40 &   0.64
     &   3.88 &   0.26 &   0.39\\
0.130-0.160
     &   2.09 &   0.59 &   0.92
     &   2.64 &   0.32 &   0.29
     &   2.55 &   0.18 &   0.24\\
0.160-0.200
     &   1.21 &   0.46 &   0.66
     &   1.64 &   0.24 &   0.29
     &   1.75 &   0.15 &   0.09\\
0.200-0.250
     &   0.49 &   0.35 &   0.45
     &   0.93 &   0.21 &   0.36
     &   0.76 &   0.13 &   0.47\\
0.250-0.300
     &   1.11 &   0.46 &   0.69
     &   0.27 &   0.20 &   0.38
     &   0.66 &   0.16 &   0.25\\
0.300-0.350
     &  -.002 &  0.345 &  0.866
     &   0.40 &   0.30 &   0.52
     &  0.058 &  0.149 &  0.219\\
        \hline
        \bw  & \multicolumn{3}{c||}{ $R(\bw)$  (172 GeV) }
        & \multicolumn{3}{c||}{ $R(\bw)$  (183 GeV) }
        & \multicolumn{3}{c||}{ $R(\bw)$  (189 GeV) } \\  
        \hline\hline
0.000-0.020
     &   8.91 &   1.75 &   3.96
     &   4.81 &   0.59 &   0.93
     &   4.61 &   0.34 &   0.91\\
0.020-0.030
     &   14.5 &    2.7 &    4.0
     &   18.1 &    1.5 &    1.0
     &   16.8 &    0.8 &    1.8\\
0.030-0.040
     &   13.7 &    2.4 &    4.5
     &   13.1 &    1.2 &    1.0
     &   13.7 &    0.7 &    0.5\\
0.040-0.050
     &   8.48 &   1.89 &   3.88
     &   10.5 &    1.0 &    0.4
     &   10.3 &    0.6 &    0.8\\
0.050-0.065
     &   7.36 &   1.49 &   2.67
     &   8.23 &   0.76 &   0.86
     &   7.89 &   0.43 &   0.42\\
0.065-0.080
     &   6.41 &   1.41 &   2.33
     &   5.16 &   0.61 &   1.06
     &   5.44 &   0.37 &   0.35\\
0.080-0.100
     &   3.88 &   1.01 &   1.43
     &   4.51 &   0.51 &   0.51
     &   4.30 &   0.30 &   0.32\\
0.100-0.150
     &   1.57 &   0.46 &   0.52
     &   2.30 &   0.26 &   0.38
     &   2.66 &   0.17 &   0.27\\
0.150-0.200
     &   1.02 &   0.43 &   0.31
     &   1.25 &   0.24 &   0.28
     &   1.03 &   0.13 &   0.24\\
0.200-0.250
     &   0.77 &   0.35 &   0.39
     &   0.33 &   0.14 &   0.21
     &   0.53 &   0.11 &   0.23\\
0.250-0.300
     &  0.048 &  0.099 &  0.331
     &  0.014 &  0.058 &  0.195
     &  0.048 &  0.038 &  0.074\\
        \hline
        \ytwothree & \multicolumn{3}{c||}{ $R(\ytwothree)$  (172 GeV) }
        & \multicolumn{3}{c||}{ $R(\ytwothree)$  (183 GeV) }
        & \multicolumn{3}{c||}{ $R(\ytwothree)$  (189 GeV) } \\  
        \hline\hline
0.00030-0.00075
     &   399. &    70. &   139.
     &   335. &    31. &    56.
     &   348. &    18. &    56.\\
0.00075-0.00130
     &   202. &    42. &    40.
     &   215. &    21. &    25.
     &   163. &    10. &    31.\\
0.00130-0.00230
     &   84.5 &   19.7 &   41.5
     &   101. &    10. &     9.
     &   111. &     6. &    10.\\
0.00230-0.00400
     &   56.4 &   12.1 &   24.6
     &   56.3 &    5.9 &    9.8
     &   62.5 &    3.6 &    7.5\\
0.00400-0.00700
     &   34.0 &    7.2 &    5.6
     &   29.7 &    3.2 &    5.1
     &   29.7 &    1.9 &    6.4\\
0.00700-0.01200
     &   16.8 &    4.2 &    6.0
     &   18.7 &    2.1 &    5.4
     &   16.6 &    1.1 &    2.6\\
0.01200-0.02300
     &   8.20 &   2.10 &   1.99
     &   7.77 &   0.96 &   1.50
     &   9.01 &   0.61 &   1.14\\
0.02300-0.04000
     &   2.56 &   0.98 &   1.74
     &   4.67 &   0.60 &   1.06
     &   4.21 &   0.35 &   0.96\\
0.04000-0.07000
     &   2.46 &   0.73 &   1.47
     &   1.74 &   0.30 &   0.62
     &   1.98 &   0.19 &   0.25\\
0.07000-0.13000
     &   0.19 &   0.24 &   1.02
     &   0.72 &   0.16 &   0.28
     &   0.76 &   0.10 &   0.08\\
0.13000-0.23500
     &   0.45 &   0.19 &   0.28
     &   0.19 &   0.07 &   0.08
     &   0.27 &   0.05 &   0.10\\
0.23500-0.40000
     &  0.036 &  0.048 &  0.074
     &  0.050 &  0.054 &  0.111
     &  0.014 &  0.018 &  0.037\\
        \hline  
      \end{tabular}
      }
  \end{center}
  \caption[] 
  { Results $R(y)=\sdsd y$ at $\rs=172$, 183 and 189~GeV
    for the event shape observables $y$: total jet broadening \bt,
    wide jet broadening \bw\ and the transition value between 2- and
    3-jets \ytwothree. The first error is statistical, the second
    systematic.  }
  \label{t:evsset3}
\end{table}

\begin{table}[!htb]
  \begin{center}
    \begin{tabular}{|r||r|r|r|r|r|r|r|}   \hline
172 GeV	&	$1-T$	&	$M_H$	&	$B_T$	&	$B_W$	&	$C$	&	\ytwothree    	&	Mean	\\ \hline 
fit range	& \small	0.05-0.30	& \small	0.17-0.45	& \small	0.075-0.25	& \small 	0.05-0.20	& \small	0.18-0.60	& \small	0.0023-0.130	&	-	\\ \hline 
$\alpha_s$	&	0.0999	&	0.0912	&	0.0908	&	0.0897	&	0.0910	&	0.0976	&	0.0919	\\
$\chi^2$/dof	&	6.7/5	&	5.0/4	&	3.3/5	&	2.7/4	&	5.9/4	&	7.9/6	&	-	\\ \hline
Stat	&	$\pm$0.0094	&	$\pm$0.0083	&	$\pm$0.0076	&	$\pm$0.0061	&	$\pm$0.0095	&	$\pm$0.0073	&	$\pm$0.0064	\\ \cline{2-8}
Exp	&	$\pm$0.0127	&	$\pm$0.0084	&	$\pm$0.0085	&	$\pm$0.0061	&	$\pm$0.0100	&	$\pm$0.0203	&	$\pm$0.0074	\\
Hadr	&	$\pm$0.0019	&	$\pm$0.0016	&	$\pm$0.0032	&	$\pm$0.0014	&	$\pm$0.0023	&	$\pm$0.0028	&	$\pm$0.0015	\\ \cline{2-8}
$\xmu=0.5$	&	$-$0.0030	&	$-$0.0015	&	$-$0.0029	&	$-$0.0018	&	$-$0.0024	&	+0.0000	&	$-$0.0020	\\
$\xmu=2$	&	+0.0038	&	+0.0023	&	+0.0036	&	+0.0025	&	+0.0031	&	+0.0016	&	+0.0028	\\ \cline{2-8}
Syst	&	$\pm$0.0133	&	$\pm$0.0087	&	$\pm$0.0096	&	$\pm$0.0066	&	$\pm$0.0106	&	$\pm$0.0205	&	$\pm$0.0079	\\ \cline{2-8}
Total	&	$\pm$0.0163	&	$\pm$0.0121	&	$\pm$0.0123	&	$\pm$0.0090	&	$\pm$0.0142	&	$\pm$0.0217	&	$\pm$0.0102	\\ \hline \hline \hline
															
183 GeV	&	$1-T$	&	$M_H$	&	$B_T$	&	$B_W$	&	$C$	&	\ytwothree 	&	Mean	\\ \hline 
fit range	& \small	0.05-0.30	& \small	0.17-0.45	& \small	0.075-0.25	& \small 	0.05-0.20	& \small	0.18-0.60	& \small	0.0023-0.130	&	-	\\ \hline 
$\alpha_s$	&	0.1070	&	0.1074	&	0.1050	&	0.1001	&	0.1088	&	0.1073	&	0.1056	\\
$\chi^2$/dof	&	3.6/5	&	2.9/4	&	6.8/5	&	2.8/4	&	1.8/4	&	3.0/6	&	-	\\ \hline
Stat	&	$\pm$0.0032	&	$\pm$0.0036	&	$\pm$0.0029	&	$\pm$0.0025	&	$\pm$0.0039	&	$\pm$0.0032	&	$\pm$0.0027	\\ \cline{2-8}
Exp	&	$\pm$0.0033	&	$\pm$0.0032	&	$\pm$0.0056	&	$\pm$0.0027	&	$\pm$0.0042	&	$\pm$0.0024	&	$\pm$0.0019	\\
Hadr	&	$\pm$0.0015	&	$\pm$0.0017	&	$\pm$0.0021	&	$\pm$0.0008	&	$\pm$0.0016	&	$\pm$0.0016	&	$\pm$0.0010	\\ \cline{2-8}
$\xmu=0.5$	&	$-$0.0038	&	$-$0.0028	&	$-$0.0045	&	$-$0.0030	&	$-$0.0043	&	$-$0.0004	&	$-$0.0026	\\
$\xmu=2$	&	+0.0048	&	+0.0040	&	+0.0055	&	+0.0038	&	+0.0054	&	+0.0024	&	+0.0039	\\ \cline{2-8}
Syst	&	$\pm$0.0056	&	$\pm$0.0050	&	$\pm$0.0078	&	$\pm$0.0044	&	$\pm$0.0066	&	$\pm$0.0032	&	$\pm$0.0039	\\ \cline{2-8}
Total	&	$\pm$0.0065	&	$\pm$0.0062	&	$\pm$0.0083	&	$\pm$0.0051	&	$\pm$0.0077	&	$\pm$0.0045	&	$\pm$0.0047	\\ \hline \hline \hline
															
189 GeV	&	$1-T$	&	$M_H$	&	$B_T$	&	$B_W$	&	$C$	&	\ytwothree 	&	Mean	\\ \hline 
fit range	& \small	0.05-0.30	& \small	0.17-0.45	& \small	0.075-0.25	& \small 	0.05-0.20	& \small	0.18-0.60	& \small	0.0023-0.130	&	-	\\ \hline 
$\alpha_s$	&	0.1122	&	0.1053	&	0.1064	&	0.1011	&	0.1086	&	0.1095	&	0.1067	\\
$\chi^2$/dof	&	2.0/5	&	5.1/4	&	7.6/5	&	5.1/4	&	1.7/4	&	6.4/6	&	-	\\ \hline
Stat	&	$\pm$0.0016	&	$\pm$0.0017	&	$\pm$0.0014	&	$\pm$0.0013	&	$\pm$0.0018	&	$\pm$0.0017	&	$\pm$0.0013	\\ \cline{2-8}
Exp	&	$\pm$0.0021	&	$\pm$0.0027	&	$\pm$0.0041	&	$\pm$0.0020	&	$\pm$0.0019	&	$\pm$0.0036	&	$\pm$0.0021	\\
Hadr	&	$\pm$0.0013	&	$\pm$0.0018	&	$\pm$0.0020	&	$\pm$0.0008	&	$\pm$0.0014	&	$\pm$0.0017	&	$\pm$0.0008	\\ \cline{2-8}
$\xmu=0.5$	&	$-$0.0045	&	$-$0.0029	&	$-$0.0047	&	$-$0.0031	&	$-$0.0044	&	$-$0.0004	&	$-$0.0030	\\
$\xmu=2$	&	+0.0056	&	+0.0040	&	+0.0056	&	+0.0040	&	+0.0054	&	+0.0026	&	+0.0043	\\ \cline{2-8}
Syst	&	$\pm$0.0056	&	$\pm$0.0047	&	$\pm$0.0069	&	$\pm$0.0041	&	$\pm$0.0054	&	$\pm$0.0043	&	$\pm$0.0043	\\ \cline{2-8}
Total	&	$\pm$0.0058	&	$\pm$0.0050	&	$\pm$0.0070	&	$\pm$0.0043	&	$\pm$0.0057	&	$\pm$0.0046	&	$\pm$0.0045	\\ \hline \hline \hline

    \end{tabular}
  \end{center}
  \caption[]
  { Values of \as(172~GeV), \as(183~GeV) and \as(189~GeV) derived using
    the \oaa+NLLA QCD calculations with $\xmu=1$ and the
    $\ln(R)$-matching scheme, fit ranges and \chisqd\ values for each
    of the six event shape observables. In addition, the statistical,
    and various systematic uncertainties are given. 
    The sign indicates
    the direction in which \as\ changes with respect to the standard analysis.
}
  \label{tab_asresults}
\end{table}

\begin{table}[!htb]
  \begin{center}
    \begin{tabular}{|r||r|r|r||r|}   \hline
         	&	172 GeV	&	183 GeV	&	189 GeV	&	187 GeV	\\	\hline\hline
$\alpha_s(Q)$	&	0.0919	&	0.1056	&	0.1067	&	0.1060	\\	\hline
Statistical error	&	$\pm$0.0064	&	$\pm$0.0027	&	$\pm$0.0013	&	$\pm$0.0012	\\	\hline\hline
Tracks+Clusters	&	$-$0.0017	&	+0.0004	&	+0.0005	&	+0.0004	\\	
Tracks Only	&	+0.0019	&	+0.0008	&	+0.0019	&	+0.0017	\\	
$|\cos\theta_T | < 0.7$	&	$-$0.0049	&	+0.0002	&	$-$0.0003	&	$-$0.0003	\\	
Alternative \rsp&	+0.0035	&	+0.0012	&	+0.0007	&	+0.0009	\\	
$\wqcd$	&	+0.0024	&	+0.0012	&	+0.0003	&	+0.0002	\\	 
Background $\pm$5\perc	&	+0.0004	&	+0.0002	&	+0.0002	&	+0.0002	\\	\hline
Experimental syst.	&	$\pm$0.0074	&	$\pm$0.0019	&	$\pm$0.0021	&	$\pm$0.0020	\\	\hline\hline
$b-1$ s.d.	&	$-$0.0001	&	$-$0.0001	&	$-$0.0001	&	$-$0.0001	\\	
$b+1$ s.d.	&	+0.0001	&	+0.0001	&	+0.0001	&	+0.0001	\\	
$\sigma_q-1$ s.d.	&	+0.0002	&	+0.0001	&	+0.0001	&	+0.0001	\\	
$\sigma_q+1$ s.d.	&	$-$0.0002	&	$-$0.0001	&	$-$0.0001	&	$-$0.0001	\\	
$Q_0=4$ GeV	&	+0.0001	&	+0.0004	&	+0.0003	&	+0.0003	\\	
udsc only	&	+0.0010	&	+0.0003	&	+0.0002	&	+0.0003	\\	
HERWIG 5.9	&	$-$0.0006	&	$-$0.0003	&	$-$0.0002	&	$-$0.0003	\\	
ARIADNE 4.08	&	+0.0009	&	+0.0007	&	+0.0007	&	+0.0007	\\	\hline
Total hadronisation	&	$\pm$0.0015	&	$\pm$0.0010	&	$\pm$0.0008	&	$\pm$0.0009	\\	\hline\hline
$\xmu=0.5$	&	$-$0.0020	&	$-$0.0026	&	$-$0.0030	&	$-$0.0029	\\	
$\xmu=2$	&	+0.0028	&	+0.0039	&	+0.0043	&	+0.0041	\\	\hline\hline
Tot. syst. error	&	$\pm$0.0079	&	$\pm$0.0039	&	$\pm$0.0043	&	$\pm$0.0041	\\	\hline

    \end{tabular}
  \end{center}
  \caption[]
  { Weighted mean values of \as, derived from fits to six event shapes
    using the \oaa+NLLA QCD calculations with $\xmu=1$ and the
    $\ln(R)$-matching scheme, for c.m.\ energies at 172, 183 and 189
    GeV.  Full detailed breakdown of the systematic uncertainties are
    listed.  The sign indicates the direction in which \as\ changes
    with respect to the standard analysis. The last column correspond
    to the values of \as, constructed at the luminosity weighted
    mean c.m.\ energy.
    }
  \label{tab_asmean}
\end{table}

\begin{table}[!htb]
  \begin{center}
    \begin{tabular}{|r||r|r|r||r|}   \hline
                        &       172 GeV &       183 GeV &       189 GeV &       187 GeV \\ \hline\hline
       $\asmz$          &       0.0992  &       0.1165  &       0.1183  &       0.1173  \\ \hline
Statistical error       &  $\pm$0.0072  &  $\pm$0.0033  &  $\pm$0.0016  &  $\pm$0.0015  \\
Total error             &  $\pm$0.0120  &  $\pm$0.0057  &  $\pm$0.0056  &  $\pm$0.0053  \\ \hline
    \end{tabular}
  \end{center}
  \caption[]
  { Weighted mean values of \as, as determined in Table~\ref{tab_asmean}, 
    evolved to \mz. 
    }
  \label{tab_asmz}
\end{table}

%
%
\begin{table}[!ht]
  \begin{center}
    {\small
    \begin{tabular}{|c| |r@{ $\pm$ }r@{ $\pm$ }r|
        r@{ $\pm$ }r@{ $\pm$ }r|
        r@{ $\pm$ }r@{ $\pm$ }r|}
      \hline
      $\nch $ &\multicolumn{3}{c|}{ P(\nch) 172 GeV $\left[ \perc \right] $}
      &\multicolumn{3}{c|}        { P(\nch) 183 GeV $\left[ \perc \right] $}
      &\multicolumn{3}{c|}        { P(\nch) 189 GeV $\left[ \perc \right] $}\\
      \hline\hline
      8.&.24&.19&.28&.09&.05&.08&.17&.03&.05\\
10.&.68&.38&.42&.35&.08&.13&.54&.06&.12\\
12.&1.31&.34&.61&1.17&.16&.18&1.34&.10&.27\\
14.&3.72&.70&.78&2.54&.26&.26&3.01&.16&.40\\
16.&5.81&.83&1.01&4.96&.36&.49&5.13&.21&.47\\
18.&8.29&1.01&1.31&7.67&.47&.69&7.15&.26&.44\\
20.&10.30&1.13&1.15&9.57&.52&.72&8.67&.28&.41\\
22.&10.12&1.02&.98&10.40&.53&.56&9.46&.28&.43\\
24.&10.32&1.05&.78&10.35&.53&.47&9.72&.29&.42\\
26.&9.10&.92&.86&9.68&.49&.51&9.44&.28&.36\\
28.&9.13&.99&.85&8.59&.45&.60&8.77&.27&.26\\
30.&7.73&.88&.76&7.30&.43&.66&7.87&.25&.25\\
32.&6.32&.79&.75&6.00&.37&.55&6.70&.23&.25\\
34.&5.01&.69&.83&4.81&.34&.37&5.51&.22&.26\\
36.&3.69&.57&.97&3.80&.32&.27&4.34&.19&.26\\
38.&2.42&.44&.86&3.13&.31&.25&3.36&.17&.30\\
40.&1.71&.37&.68&2.61&.30&.26&2.52&.15&.31\\
42.&1.19&.31&.45&2.11&.27&.21&1.88&.13&.27\\
44.&.55&.16&.26&1.61&.26&.19&1.35&.12&.24\\
46.&.36&.16&.22&1.15&.27&.14&1.00&.12&.21\\
48.&.30&.25&.26&.88&.21&.16&.72&.11&.21\\
50.&.35&.38&.30&.56&.22&.21&.50&.10&.18\\
52.&.07&.07&.29&.32&.22&.24&.34&.09&.16\\
54.&.01&.02&.28&.19&.11&.19&.21&.09&.12\\
56.&.02&.04&.31&.14&.12&.14&.12&.06&.09\\
58.&.26&.39&.25&.00&.00&.12&.06&.04&.11\\
60.&.01&.03&.72&.03&.04&.13&.09&.10&.09\\
62.&.01&.02&.28&.00&.00&.13&.02&.06&.09\\\hline\hline
\mnch&25.77&0.58&0.88&26.85&0.27&0.52&26.95&0.16&0.51\\
$D$&8.38&0.63&1.29&8.36&0.20&0.46&8.45&0.12&0.34\\
$\mnch/D$&3.08&0.21&0.45&3.21&0.07&0.14&3.19&0.04&0.09\\
\ctwo&1.106&0.014&0.030&1.097&0.004&0.017&1.098&0.003&0.005\\
\rtwo&1.067&0.014&0.031&1.060&0.004&0.017&1.061&0.003&0.006\\
      \hline
    \end{tabular}
    }
  \end{center}
  \caption[]{ 
    Charged particle multiplicity in percent, measured at $\rs=172$,  183 and 189~GeV.
    The first error is statistical, the second systematic. 
    }
  \label{tab_ch}
\end{table}


\begin{table}[!htb]
  \begin{center}
    \begin{tabular}{|r||r|r|r|}
      \hline
     & \multicolumn{3}{c |}{ \mnch} \\
     \cline{2-4} 
     &  172 GeV & 183 GeV & 189 GeV \\
      \hline\hline
      standard result &   25.77  &    26.85 &   26.95    \\ 
      \hline\hline
      statistical error & 0.58  &    0.27 &   0.16    \\
      \hline
       Alternative \rsp  &$-$0.15 &       $+$0.01  &         $-$0.01\\
Background $\pm 5$\perc  &$+$0.01 &       $+$0.02  &         $-$0.02\\
      $|d_0|<5$ cm    &$+$0.29    &       $+$0.05  &         $+$0.05\\
      $|z_0|<10$ cm   &$+$0.33    &       $+$0.35  &         $+$0.34\\
      $N_{hits}>80$   &$-$0.07    &       $-$0.08  &         $-$0.08\\
      $|\costt|<0.7$  &$-$0.05    &       $+$0.12  &         $+$0.20\\
      HERWIG          &$<$0.01    &       $-$0.35  &         $-$0.20\\
      $W_{QCD}$       &$+$0.74    &       $-$0.05  &         $+$0.23\\
      \hline
      Tot. syst. error&    0.88         & 0.52  &        0.51 \\
      \hline
    \end{tabular}
    \caption[]{ 
      Results with statistical and systematic uncertainties for
      the mean charged multiplicity \mnch\
      at $\rs =172$, 183 and 189 GeV.
      }
  \end{center}
  \label{tab_nchsyst}
\end{table}

                                %
\begin{table}[!ht]
  \begin{center}
{\small
    \begin{tabular}{|c| |r@{ $\pm$ }l@{ $\pm$ }l|
        r@{ $\pm$ }l@{ $\pm$ }l|
        r@{ $\pm$ }l@{ $\pm$ }l|}
      \hline
      $\ptin$ &\multicolumn{3}{c|}{ $R(\ptin)$ (172 GeV) }
      &\multicolumn{3}{c|}{ $R(\ptin)$ (183 GeV) }
      &\multicolumn{3}{c|}{ $R(\ptin)$ (189 GeV) }\\ 
      \hline\hline
0.0-0.1& 52.3 &  2.2 &  2.1 & 55.4 &  1.0 &  1.9 & 53.7 &  0.6 &  1.7 \\
0.1-0.2& 39.8 &  1.6 &  1.7 & 42.4 &  0.8 &  1.3 & 43.4 &  0.5 &  1.2 \\
0.2-0.3& 31.2 &  1.3 &  1.1 & 32.1 &  0.7 &  0.9 & 32.7 &  0.4 &  0.7 \\
0.3-0.4& 23.9 &  1.1 &  1.0 & 24.2 &  0.6 &  0.6 & 24.1 &  0.3 &  0.4 \\
0.4-0.5& 18.6 &  1.1 &  0.7 & 19.1 &  0.5 &  0.5 & 17.9 &  0.3 &  0.3 \\
0.5-0.6& 16.3 &  1.0 &  0.6 & 13.7 &  0.4 &  0.4 & 14.8 &  0.3 &  0.3 \\
0.6-0.7& 10.9 &  0.9 &  0.4 & 11.4 &  0.4 &  0.3 & 11.1 &  0.2 &  0.2 \\
0.7-0.8&  8.43&  0.70&  0.37&  9.40&  0.36&  0.21&  8.65&  0.20&  0.17\\
0.8-0.9&  7.62&  0.74&  0.40&  7.09&  0.31&  0.18&  7.46&  0.19&  0.13\\
0.9-1.0&  5.67&  0.60&  0.38&  5.92&  0.30&  0.22&  6.42&  0.18&  0.19\\
1.0-1.2&  3.88&  0.39&  0.40&  4.52&  0.19&  0.23&  4.95&  0.12&  0.16\\
1.2-1.4&  2.82&  0.35&  0.39&  3.95&  0.19&  0.25&  3.56&  0.10&  0.13\\
1.4-1.6&  2.28&  0.31&  0.31&  2.65&  0.15&  0.17&  2.94&  0.09&  0.08\\
1.6-2.0&  1.78&  0.22&  0.19&  1.89&  0.10&  0.12&  1.95&  0.06&  0.06\\
2.0-2.5&  1.00&  0.14&  0.11&  1.10&  0.07&  0.06&  1.16&  0.04&  0.05\\
2.5-3.0&  0.47&  0.09&  0.06&  0.75&  0.06&  0.05&  0.72&  0.03&  0.03\\
3.0-3.5&  0.38&  0.09&  0.07&  0.49&  0.04&  0.05&  0.55&  0.03&  0.03\\
3.5-4.0&  0.27&  0.08&  0.06&  0.29&  0.04&  0.03&  0.37&  0.02&  0.02\\
4.0-5.0&  0.17&  0.04&  0.05&  0.22&  0.02&  0.03&  0.23&  0.01&  0.02\\
5.0-6.0&  0.06&  0.03&  0.02&  0.13&  0.02&  0.02&  0.12&  0.01&  0.01\\
6.0-7.0&  0.07&  0.03&  0.03&  0.05&  0.01&  0.02&  0.08&  0.01&  0.01\\
7.0-8.0&  0.06&  0.03&  0.03&  0.02&  0.01&  0.01&  0.04&  0.01&  0.01\\
\hline
$\langle \ptin \rangle$& 0.593& 0.018& 0.016& 0.622& 0.008& 0.009& 0.647& 0.005& 0.011\\

      \hline
    \end{tabular}
}
  \end{center}
  \caption[]{ 
    Measured values for the momentum spectra $R(\ptin)=\sdscd\ptin$ in
    the event plane at $\rs =172$, 183 and 189 GeV.  The
    first error is statistical, the second systematic. The mean values
    are also shown.  }
  \label{tab_ptin}
\end{table}

%
\begin{table}[!ht]
  \begin{center}
{\small
    \begin{tabular}{|c| |r@{ $\pm$ }l@{ $\pm$ }l|
        r@{ $\pm$ }l@{ $\pm$ }l|
        r@{ $\pm$ }l@{ $\pm$ }l|}
      \hline
      $\ptout$ &\multicolumn{3}{c|}{ $R(\ptout)$ (172 GeV) }
      &\multicolumn{3}{c|}{ $R(\ptout)$ (183 GeV) }
      &\multicolumn{3}{c|}{ $R(\ptout)$ (189 GeV) }\\ 
      \hline\hline
0.0-0.1& 71.5 &  2.3 &  2.3 & 77.2 &  1.2 &  1.7 & 76.2 &  0.7 &  1.6 \\
0.1-0.2& 59.1 &  2.0 &  1.7 & 59.1 &  1.0 &  1.1 & 59.8 &  0.6 &  1.1 \\
0.2-0.3& 39.7 &  1.6 &  1.1 & 41.1 &  0.7 &  0.6 & 41.7 &  0.5 &  0.7 \\
0.3-0.4& 26.2 &  1.3 &  0.9 & 27.9 &  0.6 &  0.5 & 27.6 &  0.4 &  0.4 \\
0.4-0.5& 18.4 &  1.2 &  0.7 & 19.0 &  0.5 &  0.5 & 18.8 &  0.3 &  0.4 \\
0.5-0.6& 11.8 &  1.0 &  0.5 & 11.6 &  0.4 &  0.3 & 12.2 &  0.3 &  0.3 \\

0.6-0.7&  7.63&  0.75&  0.51&  8.29&  0.36&  0.29&  8.45&  0.22&  0.27\\
0.7-0.8&  3.94&  0.57&  0.41&  5.30&  0.29&  0.19&  5.65&  0.18&  0.19\\
0.8-0.9&  3.56&  0.49&  0.59&  4.15&  0.27&  0.25&  4.11&  0.16&  0.13\\
0.9-1.0&  2.46&  0.45&  0.49&  2.82&  0.22&  0.20&  2.91&  0.13&  0.10\\
1.0-1.2&  1.76&  0.31&  0.36&  1.96&  0.15&  0.16&  2.04&  0.09&  0.09\\
1.2-1.4&  0.97&  0.20&  0.18&  1.25&  0.12&  0.19&  1.23&  0.07&  0.08\\
1.4-1.6&  0.62&  0.18&  0.13&  0.62&  0.09&  0.13&  0.69&  0.05&  0.06\\
1.6-2.0&  0.33&  0.11&  0.90&  0.40&  0.05&  0.13&  0.43&  0.04&  0.05\\
2.0-2.4&  0.01&  0.04&  0.03&  0.16&  0.04&  0.06&  0.20&  0.03&  0.03\\
2.4-2.8&  0.04&  0.05&  0.15&  0.04&  0.03&  0.03&  0.11&  0.02&  0.03\\
2.8-3.2& \multicolumn{3}{c|}{ \--- }& 0.04&  0.03&  0.10&  0.07&  0.02&  0.04\\
3.2-3.6& \multicolumn{3}{c|}{ \--- }& 0.01&  0.02&  0.01&  0.04&  0.02&  0.05\\
3.6-4.0& \multicolumn{3}{c|}{ \--- }& \multicolumn{3}{c|}{ \--- } &  0.02&  0.02&  0.03\\
\hline
$\langle \ptout \rangle$& 0.282& 0.007& 0.022& 0.289& 0.003& 0.008& 0.300& 0.003& 0.004\\

      \hline
    \end{tabular}
}
  \end{center}
  \caption[]{ 
    Measured values for the momentum spectra $R(\ptout)=\sdscd\ptout$
    out of the event plane at $\rs =172$, 183 and 189 GeV.
    The first error is statistical, the second systematic. The mean
    values are also shown.}
  \label{tab_ptout}
\end{table}

%
\begin{table}[!ht]
\begin{center}
\small{
\begin{tabular}{|c| |r@{ $\pm$ }l@{ $\pm$ }l|
                     r@{ $\pm$ }l@{ $\pm$ }l|
                     r@{ $\pm$ }l@{ $\pm$ }l|}
\hline
 $y$ &\multicolumn{3}{c|}{ $R(\yp)$ (172 GeV) }
     &\multicolumn{3}{c|}{ $R(\yp)$ (183 GeV) }
     &\multicolumn{3}{c|}{ $R(\yp)$ (189 GeV) }\\ 
\hline\hline
0.00-0.33 & 7.56 & 0.60 & 0.51 & 7.40 & 0.29 & 0.35 & 7.44 & 0.18 & 0.30 \\
0.33-0.67 & 6.86 & 0.60 & 0.41 & 7.45 & 0.29 & 0.32 & 8.08 & 0.18 & 0.29 \\
0.67-1.00 & 6.99 & 0.57 & 0.44 & 7.60 & 0.28 & 0.26 & 8.31 & 0.17 & 0.25 \\
1.00-1.33 & 6.98 & 0.49 & 0.35 & 7.51 & 0.25 & 0.23 & 7.95 & 0.15 & 0.19 \\
1.33-1.67 & 6.47 & 0.47 & 0.30 & 8.07 & 0.24 & 0.22 & 7.79 & 0.14 & 0.19 \\
1.67-2.00 & 7.74 & 0.43 & 0.34 & 7.30 & 0.21 & 0.19 & 7.36 & 0.12 & 0.16 \\
2.00-2.33 & 6.72 & 0.41 & 0.28 & 6.99 & 0.19 & 0.21 & 7.02 & 0.11 & 0.14 \\
2.33-2.67 & 6.40 & 0.36 & 0.27 & 6.45 & 0.17 & 0.18 & 6.40 & 0.10 & 0.12 \\
2.67-3.00 & 5.85 & 0.34 & 0.24 & 5.89 & 0.15 & 0.16 & 5.77 & 0.09 & 0.13 \\
3.00-3.33 & 4.45 & 0.28 & 0.20 & 4.96 & 0.14 & 0.13 & 4.91 & 0.08 & 0.14 \\
3.33-3.67 & 4.37 & 0.32 & 0.20 & 4.00 & 0.13 & 0.13 & 3.80 & 0.07 & 0.11 \\
3.67-4.00 & 2.38 & 0.21 & 0.11 & 2.97 & 0.11 & 0.12 & 2.72 & 0.06 & 0.08 \\
4.00-4.33 & 1.62 & 0.16 & 0.11 & 1.76 & 0.09 & 0.08 & 1.77 & 0.05 & 0.07 \\
4.33-4.67 & 1.19 & 0.15 & 0.12 & 1.00 & 0.06 & 0.07 & 0.89 & 0.03 & 0.07 \\
4.67-5.00 & 0.31 & 0.08 & 0.05 & 0.51 & 0.05 & 0.05 & 0.50 & 0.03 & 0.05 \\
5.00-5.33 & 0.19 & 0.06 & 0.06 & 0.22 & 0.03 & 0.03 & 0.21 & 0.02 & 0.02 \\
5.33-5.67 & 0.045& 0.026& 0.028& 0.089& 0.017& 0.015& 0.082& 0.010& 0.010\\
5.67-6.00 & 0.019& 0.019& 0.017& 0.039& 0.011& 0.009& 0.031& 0.006& 0.006\\
6.00-6.33 & 0.012& 0.012& 0.013& 0.007& 0.004& 0.001& 0.007& 0.003& 0.002\\
\hline
$\langle y \rangle$& 1.877& 0.024& 0.032& 1.877& 0.011& 0.026& 1.837& 0.006& 0.018\\

\hline
\end{tabular}
}
\end{center}
\caption[]{ Measured values for the rapidity $R(\yp)=\sdscd\yp$ distributions
 at $\rs =172$, 183 and 189 GeV.
The first error is statistical, the second systematic. The mean values are also
shown.}
\label{tab_Y}
\end{table}

\begin{table}[!ht]
\begin{center}
{\small
\begin{tabular}{|c|r@{ $\pm$ }l@{ $\pm$ }l|
                   r@{ $\pm$ }l@{ $\pm$ }l|
                   r@{ $\pm$ }l@{ $\pm$ }l|} 
\hline
  \xp & \multicolumn{3}{c||}{ $R(\xp)$  (172 GeV) }
      & \multicolumn{3}{c||}{ $R(\xp)$  (183 GeV) }
      & \multicolumn{3}{c||}{ $R(\xp)$  (189 GeV) } \\  
\hline\hline
0.00-0.01&749.  & 30.  & 30.  &812.  & 15.  & 21.  &855.  &  9.  & 20.  \\
0.01-0.02&472.  & 20.  & 22.  &515.  & 10.  & 13.  &506.  &  6.  & 10.  \\
0.02-0.03&281.  & 15.  & 11.  &295.  &  7.  &  7.  &295.  &  4.  &  7.  \\
0.03-0.04&194.  & 12.  &  8.  &191.  &  5.  &  6.  &192.  &  3.  &  4.  \\
0.04-0.05&130.  & 10.  &  6.  &147.  &  4.  &  4.  &140.  &  3.  &  3.  \\
0.05-0.06&117.  &  8.  &  7.  &103.  &  4.  &  3.  &106.  &  2.  &  2.  \\
0.06-0.07& 77.4 &  7.3 &  4.3 & 87.9 &  3.4 &  2.3 & 84.4 &  2.0 &  2.6 \\
0.07-0.08& 66.8 &  6.0 &  4.0 & 64.3 &  2.9 &  1.6 & 66.2 &  1.7 &  2.0 \\
0.08-0.09& 50.8 &  5.4 &  3.4 & 55.1 &  2.6 &  1.6 & 56.0 &  1.6 &  1.9 \\
0.09-0.10& 42.9 &  4.9 &  3.2 & 46.2 &  2.3 &  1.4 & 46.4 &  1.4 &  1.2 \\
0.10-0.12& 38.4 &  3.6 &  2.8 & 35.2 &  1.5 &  1.3 & 37.0 &  0.9 &  1.3 \\
0.12-0.14& 26.5 &  3.1 &  1.5 & 30.1 &  1.4 &  1.0 & 27.3 &  0.8 &  1.4 \\
0.14-0.16& 25.9 &  2.5 &  1.5 & 22.7 &  1.2 &  1.1 & 21.1 &  0.7 &  1.1 \\
0.16-0.18& 20.0 &  2.5 &  1.3 & 18.9 &  1.1 &  1.3 & 16.0 &  0.6 &  0.9 \\
0.18-0.20& 14.7 &  2.0 &  1.1 & 11.7 &  0.9 &  0.9 & 12.1 &  0.5 &  0.6 \\
0.20-0.25&  8.54&  0.93&  1.11&  9.66&  0.50&  0.67&  9.03&  0.28&  0.38\\
0.25-0.30&  3.92&  0.70&  0.55&  5.53&  0.37&  0.38&  5.20&  0.21&  0.22\\
0.30-0.40&  2.69&  0.34&  0.37&  2.65&  0.18&  0.16&  2.60&  0.10&  0.10\\
0.40-0.50&  1.09&  0.24&  0.14&  1.34&  0.12&  0.14&  1.13&  0.07&  0.06\\
0.50-0.60&  0.77&  0.18&  0.12&  0.51&  0.08&  0.08&  0.50&  0.04&  0.04\\
0.60-0.80&  0.16&  0.06&  0.06&  0.12&  0.02&  0.04&  0.14&  0.02&  0.04\\
\hline
$\langle \xp \rangle$&0.0495&0.0011&0.0010&0.0480&0.0005&0.0007&0.0464&0.0003&0.0007\\

\hline  
\end{tabular}
}
\end{center}
\caption[]{ Measured values for the fragmentation 
functions $R(\xp)=\sdscd\xp$ 
 at $\rs =172$, 183 and 189 GeV. 
 The first error is statistical, the second systematic. The mean values are also
shown.}
\label{tab_xp}
\end{table}

\begin{table}[!ht]
\begin{center}
  {\small
\begin{tabular}{|c|r@{ $\pm$ }l@{ $\pm$ }l|
                   r@{ $\pm$ }l@{ $\pm$ }l|
                   r@{ $\pm$ }l@{ $\pm$ }l|} 
\hline
  \ksip &\multicolumn{3}{c||}{ $R(\ksip)$ (172 GeV)} 
        &\multicolumn{3}{c||}{ $R(\ksip)$ (183 GeV)} 
        &\multicolumn{3}{c||}{ $R(\ksip)$ (189 GeV)}  \\    
\hline\hline
 0.0-0.2& 0.005& 0.006& 0.003& 0.015& 0.005& 0.005& 0.015& 0.003& 0.005\\
 0.2-0.4&  0.05&  0.03&  0.02&  0.07&  0.02&  0.02&  0.08&  0.01&  0.02\\
 0.4-0.6&  0.30&  0.08&  0.06&  0.11&  0.03&  0.02&  0.16&  0.02&  0.02\\
 0.6-0.8&  0.42&  0.10&  0.06&  0.46&  0.05&  0.09&  0.36&  0.03&  0.04\\
 0.8-1.0&  0.57&  0.12&  0.07&  0.65&  0.06&  0.06&  0.63&  0.03&  0.05\\
 1.0-1.2&  1.01&  0.15&  0.14&  1.00&  0.08&  0.08&  0.96&  0.04&  0.05\\
 1.2-1.4&  1.12&  0.19&  0.14&  1.54&  0.09&  0.10&  1.42&  0.05&  0.06\\
 1.4-1.6&  1.90&  0.22&  0.24&  2.07&  0.12&  0.15&  1.98&  0.06&  0.08\\
 1.6-1.8&  2.86&  0.31&  0.20&  2.74&  0.13&  0.15&  2.46&  0.07&  0.12\\
 1.8-2.0&  3.90&  0.31&  0.19&  3.36&  0.14&  0.16&  3.18&  0.08&  0.13\\
 2.0-2.2&  3.62&  0.33&  0.20&  3.78&  0.16&  0.13&  3.61&  0.09&  0.15\\
 2.2-2.4&  4.18&  0.36&  0.24&  4.19&  0.16&  0.14&  4.33&  0.10&  0.14\\
 2.4-2.6&  4.62&  0.37&  0.28&  4.90&  0.18&  0.16&  4.82&  0.10&  0.13\\
 2.6-2.8&  5.06&  0.42&  0.26&  5.27&  0.19&  0.14&  5.40&  0.11&  0.13\\
 2.8-3.0&  6.17&  0.41&  0.32&  5.66&  0.19&  0.19&  5.76&  0.11&  0.11\\
 3.0-3.2&  5.98&  0.47&  0.24&  6.59&  0.20&  0.19&  6.25&  0.12&  0.15\\
 3.2-3.4&  5.93&  0.43&  0.23&  6.42&  0.21&  0.23&  6.66&  0.12&  0.16\\
 3.4-3.6&  8.06&  0.52&  0.36&  6.93&  0.22&  0.22&  6.89&  0.13&  0.17\\
 3.6-3.8&  6.88&  0.48&  0.44&  7.46&  0.23&  0.23&  7.31&  0.13&  0.14\\
 3.8-4.0&  6.30&  0.48&  0.42&  7.67&  0.24&  0.21&  7.41&  0.14&  0.18\\
 4.0-4.2&  7.30&  0.50&  0.51&  7.51&  0.22&  0.24&  7.13&  0.13&  0.19\\
 4.2-4.4&  6.33&  0.44&  0.37&  7.61&  0.24&  0.27&  7.45&  0.14&  0.22\\
 4.4-4.6&  6.93&  0.45&  0.41&  6.99&  0.22&  0.22&  7.28&  0.14&  0.17\\
 4.6-4.8&  6.56&  0.46&  0.45&  6.71&  0.21&  0.17&  6.90&  0.13&  0.18\\
 4.8-5.0&  6.39&  0.48&  0.44&  6.25&  0.20&  0.18&  6.48&  0.13&  0.20\\
 5.0-5.2&  5.18&  0.41&  0.42&  5.92&  0.20&  0.24&  5.89&  0.12&  0.20\\
 5.2-5.4&  4.40&  0.38&  0.37&  4.95&  0.19&  0.21&  5.32&  0.11&  0.19\\
 5.4-5.6&  3.95&  0.38&  0.40&  4.11&  0.17&  0.22&  4.49&  0.10&  0.17\\
 5.6-5.8&  2.85&  0.29&  0.24&  3.63&  0.16&  0.20&  3.74&  0.09&  0.20\\
 5.8-6.0&  2.58&  0.30&  0.29&  2.59&  0.14&  0.20&  3.02&  0.09&  0.17\\
 6.0-6.2&  2.12&  0.29&  0.29&  2.17&  0.13&  0.17&  2.34&  0.08&  0.17\\
 6.2-6.4&  0.92&  0.29&  0.43&  1.47&  0.14&  0.28&  1.62&  0.08&  0.19\\

\hline 
\end{tabular}
}
\end{center}
\caption[]{ Measured values for the $R(\ksip)=\sdscd\ksip$ distributions
 at $\rs =172$, 183 and 189 GeV.
The first error is statistical, the second systematic.}
\label{ksip_tab}
\end{table}


\begin{table}[!htb]
  \begin{center}
    \begin{tabular}{|r||r|r|r|}
      \hline
     & \multicolumn{3}{c |}{ \ksinul} \\
     \cline{2-4} 
     &  172 GeV & 183 GeV & 189 GeV \\
      \hline\hline
      standard result &   4.031  &    4.087 &   4.124    \\ 
      \hline\hline
      statistical error & 0.033  &    0.014 &   0.010    \\
      \hline
            Alternative \rsp      &    0.012 & $<$0.001 &    0.010   \\
Background $\pm$5 \perc      & $-$0.002 & $-$0.002 &    0.002   \\
             $|d_0|<5$ cm    & $-$0.006 &    0.011 & $<$0.001  \\
             $|z_0|<10$ cm   & $-$0.006 & $-$0.006 & $-$0.005  \\
             $N_{hits}>80$   & $-$0.026 & $-$0.023 & $-$0.024  \\
             $|\costt|<0.7$  &    0.012 & $-$0.002 &    0.002    \\
             HERWIG          &    0.005 &    0.013 &    0.006  \\
             $W_{QCD}$       &    0.001 & $-$0.001 & $-$0.004   \\
             fit range       &    0.024 &    0.007 &    0.025   \\
      \hline
             Tot. syst. error&    0.041 &    0.030 &    0.037    \\
      \hline
    \end{tabular}
    \caption[]{ Results with statistical and systematic uncertainties for
      the position \ksinul\ of the peak of the \ksip\ 
      distribution at $\rs =172$, 183 and 189 GeV.}
  \end{center}
\end{table}


\begin{table}[!htb]
  \begin{center}
    \begin{tabular}{|r||r|r|r|}
      \hline
     & \multicolumn{3}{c |}{  $K$} \\
     \cline{2-4} 
     &  172 GeV & 183 GeV & 189 GeV \\
      \hline\hline
      standard result &   1.143   &   1.183 &   1.164   \\ 
      \hline\hline
      statistical error & 0.030  &    0.010 &   0.008    \\
      \hline
              Alternative \rsp        &    0.014 & $-$0.001 & $<$0.001   \\
Background $\pm$5 \perc      & $-$0.001 &    0.001 & $-$0.001   \\
             $|d_0|<5$ cm    & $-$0.013 & $-$0.002 & $<$0.001  \\
             $|z_0|<10$ cm   & $-$0.012 & $-$0.012 & $-$0.012  \\
             $N_{hits}>80$   & $-$0.001 & $<$0.001 & $<$0.001  \\
             $|\costt|<0.7$  & $-$0.009 & $-$0.006 & $-$0.005    \\
             HERWIG          &    0.012 &    0.011 &    0.005  \\
             $W_{QCD}$       &    0.005 & $-$0.003 & $-$0.007   \\
             fit range       &    0.049 &    0.015 &    0.026   \\
      \hline
             Tot. syst. error&    0.056 &    0.023 &    0.030    \\
      \hline
    \end{tabular}
    \caption[]{ Results for the normalisation factor $K$ in the MLLA description of
     the \ksip\ distribution with statistical and systematic uncertainties at $\rs =172$, 
     183 and 189 GeV.}
  \end{center}
\end{table}


\begin{table}[!htb]
  \begin{center}
    \begin{tabular}{|r||r|r|r|}
      \hline
     & \multicolumn{3}{c |}{  $\ksinul-\ksimean$} \\
     \cline{2-4} 
     &  172 GeV & 183 GeV & 189 GeV \\
      \hline\hline
      standard result &   0.054   &   0.012 &  $-$0.035    \\ 
      \hline\hline
      statistical error & 0.023  &    0.010 &   0.007    \\
      \hline
            Alternative \rsp         &    0.005 & $-$0.001 &    0.008   \\
Background $\pm$5 \perc      &    0.001 &    0.002 & $-$0.003   \\
             $|d_0|<5$ cm    &    0.003 &    0.001 &    0.004  \\
             $|z_0|<10$ cm   & $-$0.002 & $-$0.002 & $-$0.003  \\
             $N_{hits}>80$   & $-$0.015 & $-$0.015 & $-$0.018  \\
             $|\costt|<0.7$  & $-$0.017 &    0.002 &    0.006    \\
             HERWIG          & $-$0.046 & $-$0.059 & $-$0.069  \\
             $W_{QCD}$       &    0.016 &    0.005 & $-$0.009   \\
             fit range       &    0.032 &    0.020 &    0.020   \\
      \hline
             Tot. syst. error&    0.063 &    0.064 &    0.075    \\
      \hline
    \end{tabular}
    \caption[]{ Results with statistical and systematic uncertainties for
      the difference of the position \ksinul\ of the peak of the \ksip\ 
      distribution and its mean value \ksimean at $\rs =172$, 183 and 189 GeV.}
  \end{center}
\end{table}


\begin{table}[!htb]
  \begin{center}
    \begin{tabular}{|r||r|r|r|}
      \hline
     & \multicolumn{3}{c |}{ \large $\frac{\ksinul-\ksimean}{\ksimed-\ksimean}$ } \\
     \cline{2-4} 
     &  172 GeV & 183 GeV & 189 GeV \\
      \hline\hline
      standard result &  $-$0.38   & $-$0.06 & 0.16    \\ 
      \hline\hline
      statistical error & 0.56  &    0.44  &   0.33    \\
      \hline
             Alternative \rsp       & $-$0.05 &    0.01 & $-$0.03   \\
Background $\pm$5 \perc      & $<$0.01 & $-$0.01 &    0.01   \\
             $|d_0|<5$ cm    & $-$0.03 & $-$0.01 & $-$0.02  \\
             $|z_0|<10$ cm   &    0.02 &    0.01 &    0.01  \\
             $N_{hits}>80$   &    0.12 &    0.08 &    0.07  \\
             $|\costt|<0.7$  &    0.15 & $-$0.01 & $-$0.02    \\
             HERWIG          &    0.34 &    0.26 &    0.20  \\
             $W_{QCD}$       & $-$0.14 & $-$0.03 &    0.04   \\
             fit range       &    0.03 &    0.02 &    0.02   \\
      \hline
             Tot. syst. error&    0.42 &    0.27 &    0.22    \\
      \hline
    \end{tabular}
    \caption[]{ Results with statistical and systematic uncertainties for
      the ratio of the difference of the position \ksinul\ of the peak of the \ksip\ 
      distribution and its mean value \ksimean and the difference between the
      median and the mean of the \ksip\ distribution, $(\ksinul-\ksimean)/(\ksimed-\ksimean)$ 
      at $\rs =172$, 183 and 189 GeV.}
  \end{center}
\end{table}

\clearpage

\end{document}